2.

# INEQUALITY AT RISK OF AUTOMATION? GENDER DIFFERENCES IN ROUTINE TASKS INTENSITY IN DEVELOPING COUNTRY LABOR MARKETS

Janneke Pieters
Ana Kujundzic
Rulof Burger
Joel Gondwe

Technological change can have profound impacts on the labor market. Decades of research have made it clear that technological change produces winners and losers. Machines can replace some types of work that humans do, while new technologies increase human's productivity in other types of work. For a long time, highly educated workers benefitted from increased demand for their labor due to skill-biased technological change, while the losers were concentrated at the bottom of the wage distribution Katz and Autor, 1999; Goldin and Katz, 2007, 2010; Kijima, 2006. Currently, however, labor markets seem to be affected by a different type of technological change, the so-called routine-biased technological change (RBTC).

RBTC is a process whereby new technologies are used to automate routine tasks that were otherwise done by human workers (e.g., Katz and Autor, 1999; Goldin and Katz, 2007, 2010; Kijima, 2006). Because different jobs consist of different combinations of tasks, some workers are more exposed to the "risk of automation" than others (e.g., Frey and Osborne, 2017). In particular, RBTC is expected to reduce labor demand in jobs with a high routine task intensity.

This chapter studies the risk of automation in developing country labor markets, with a particular focus on differences between men and women. Given the pervasiveness of gender occupational segregation, there may be important gender differences in the risk of automation. Understanding these differences is important to ensure progress towards equitable development and gender inclusion in the face of new technological advances. Our objective is to describe the gender gap in the routine task intensity of jobs in developing countries and to explore the role of occupational segregation and several worker characteristics in accounting for the gender gap.

We use individual-level harmonized survey data representative of urban labor markets in 13 low- and middle-income countries to document that the lowest-paid occupations are most routine intensive, and that women's jobs are more routine task intensive than men's. Women report higher routine task intensity than men within the same 1-digit occupational groups. We further show that gender differences in occupational choice across 2-digit occupations, as well as differences in human capital and ethnicity, account for just a small part of the gender gap in routine task intensity. These findings contribute to an understanding of gender inequalities in developing country labor markets, and how this relates to the potential impact of automation technologies. While there may currently be little incentive for employers to invest in the automation of routine tasks, given the low cost of labor, the continued decline in the cost of automation may lead to increased automation in the future, in which case the impact would be concentrated among women and low-wage workers.

## GENDER AND ROUTINE BIASED TECHNOLOGICAL CHANGE

Over the past decades, many high-income countries have experienced a process of job polarization – employment growth has been concentrated in occupations with low and high wages, while occupations in the middle of the wage distribution saw slower employment growth. Evidence from various studies indicates this phenomenon can be explained by RBTC, whereby new technologies are used to automate routine tasks, which no longer need to be performed by workers (Autor et al., 2003; Acemoglu and Autor, 2011; Goos et al., 2014).[1] Because jobs in the middle of the wage distribution tend to consist of a relatively high share of routine tasks, RBTC reduces the demand for workers in middle-skilled jobs.[2]

[1] The research on job polarization builds on the so-called task-based approach (Autor et al., 2003), which considers occupations as a collection of tasks that can be classified into routine and non-routine tasks. As Autor (2013) describes, an important advantage of the approach is that we can focus on a relatively limited set of tasks to describe the nature of work across many hundreds (or thousands) of occupations.
[2] These are jobs that pay around the median wage such as clerical and accounting jobs, plant and machine operators, and other related repetitive-motion middle-skilled occupations.

In the same way, new technologies can lead to substantial changes in developing countries' occupational structure. The empirical evidence for job polarization in developing countries is limited,[3] which could reflect differences in the occupational structure in developing countries, compared to high-income countries. Das and Hilgenstock (2018) analyze routine task intensity for 85 developed and developing country labor markets since 1960. They find developing country workers are less exposed to routinization, reflecting the low relative price of labor and the concentration of employment in manual in-person tasks. Still, exposure has increased since the 1990s due to structural change and globalization.

Like many other studies, Das and Hilgenstock (2018) use a Routine Task Intensity (RTI) measure constructed by Autor and Dorn (2013) based on US data describing the task content of occupations in the US economy.[4] There are two important drawbacks to this approach. First, the task content of jobs in developing countries may differ from that in the US – for example due to differences in the costs and availability of non-labor inputs. Using recently collected data on skills and work tasks across a range of countries (the same data used in this chapter), Dicarlo et al., (2016) show that the skill content of occupations is similar across developing countries, but differs between developing countries and the US. Lewandowski et al., (2019) compare task intensity measures based on survey data from 42 developed and developing countries and find sizable cross-country differences in task content, even within the same occupational group. Lewandowski et al., (2020) further show that within the same occupations, jobs in low- and middle-income countries are more routine intensive than in high-income countries. This implies that for analysis of developing country labor markets, using US-based data on the task content of occupations may lead to distorted results.

[3]Job polarization has been documented for Brazil, Colombia and Mexico during the early 2000s (Almeida et al., 2017; Ariza and Bara, 2020), with evidence of job polarization being restricted to only a subset of countries. However, cross-country studies have produced mixed findings (Fu et al., 2021; Longmuir et al., 2020), with evidence of job polarization being restricted to only a subset of countries.
[4] These are the US Department of Labor's Dictionary of Occupational Titles.

Second, occupation-level RTI measures mask considerable variation in task intensity across workers within the same occupation (Arnzt et al., 2017). Since within-occupation gender differences in jobs are potentially important for understanding gender gaps in routine intensity (as documented for Germany by Black and Spitz-Oener, 2010), individual-level job task measures are important to arrive at an accurate picture of gender differences in the risk of automation.

To date, only a handful of studies have looked at differences in job tasks between male and female workers. Black and Spitz-Oener (2008, 2010) investigate the implications of task polarization for German men and women. In the 1970s, women were over-represented in occupations that intensively involved routine tasks. In the decades that followed, women experienced larger reductions in their jobs' routine task content compared to men. This led to greater job polarization for women and at the same time accounted for a substantial part of the closing of the gender wage gap during the 1980s and 1990s (Black and Spitz-Oener, 2010).

Brussevich et al., (2019) analyze individual level job tasks for 30 advanced and emerging economies. They document that women's jobs are more routine task intensive compared to men's, on average, and that the gender gap in routine intensity is negatively correlated with female labor force participation, while it is positively correlated with the manufacturing share of GDP. Furthermore, they find that women's routine intensity exceeds men's within each 2-digit ISCO occupation. Looking at changes over the period 1994-2016, they note that women have disproportionately moved out of clerical and elementary occupations towards services and professional jobs. While women have thus increasingly selected into low-routine jobs, they are still more exposed to the risk of automation. Cortes and Pan (2019) reach a similar conclusion from US census and survey data for the period 1980-2017, using 3-digit occupation-level task measures from Autor and Dorn (2013). Changes in the occupational structure of men and women contributed substantially to a closing of the gender gap in routine intensity, partly because women raised their educational profile. It seems women were better able to adapt to automation-related changes in the labor market, although it remains unclear to what extent automation (as opposed to changes in secular demand, norms, and other factors) is responsible for the observed occupational shifts.

## MEASURING ROUTINE TASK INTENSITY

The analyses in this chapter are based on the World Bank's Skills Toward Employment and Productivity (STEP) data, an initiative to measure specific work tasks in low- and middle-income countries. The STEP project includes household-based surveys and employer-based surveys to assess both the supply of and demand for occupational skills. The surveys have been implemented in 18 countries so far. We analyze the 13 countries for which household survey data was collected (one cross-sectional survey in each country) between 2012 and 2017: Armenia, Bolivia, Colombia, Georgia, Ghana, Kenya, North Macedonia, Philippines, Sri Lanka, Ukraine, Vietnam, and Yunnan province of China.[5] The survey's target population consists of non-institutionalized adults 15 to 64 years of age living in private dwellings in urban areas.[6] The household surveys collect background information of all household members age six and older and more detailed information, including employment history, skills, and occupational tasks, for one individual respondent who is randomly selected among all adult household members. Individuals who were unemployed or working in armed forces occupations in the year preceding the survey are excluded from our sample of workers used in the analysis.[7]

[7] Self-employed and unpaid family workers are included.
[6] Sample sizes range from 2,989 observations in Sri Lanka to 4,009 observations in Macedonia (see Pierre et al., 2014 for technical details on the STEP surveys).
[5] The remaining five countries, where only the employer-based survey was conducted, are not included in the analysis (Albania, Azerbaijan, Bosnia & Herzegovina, Kosovo, and Serbia).

To measure the risk of automation for men and women, we construct an RTI index. As previously stated, our methodology builds on the task-based framework pioneered by Autor et al., (2003), where jobs are classified according to their task requirements and the set of skills required to accomplish these tasks. Since the original RTI measure was created to describe the task content of occupations in the US economy, we first selected the appropriate STEP survey items that best capture the five US Dictionary of Occupational Titles (DOT) task measures used in Autor et al. (2003). We follow the approach of Lo Bello et al. (2019) but with some adjustments. The mapping of survey items to three task categories (abstract, routine, and manual) is summarized in Table 2.1.[8]

The STEP task variables are measured at different scales; so, to construct the composite RTI index, we standardize each variable using sampling weights to have a mean of zero and a unit standard deviation. The standardized variables within each task category are then summed, and the sum is again standardized to obtain three task indexes that vary at the individual worker level. For example, the individual-level task index for the abstract category is the standardized sum of five standardized variables ("Thinking at work," "Learning at work," "Contact with clients/suppliers," "Formal presentation to clients," and "Supervising co-workers"). Standardization is always done within countries since we analyze each country separately in our subsequent analyses.

The RTI index is calculated as:

$$RTI = R - (A + M) \qquad (1)$$

where R, A, and M are the Routine, Abstract, and Manual task indexes. The RTI index varies at the individual worker level and is increasing in R and decreasing in A and M. In other words, the higher the value of RTI, the more routine intense the job is. To obtain the occupational-level RTI index, we calculate the average of the individual-level RTI indexes for each 1-digit ISCO-08 occupational group using sampling weights.

[8] Autor et al., (2003) map DOT task variables onto five task categories: non-routine analytical, non-routine interpersonal, routine cognitive, routine manual, and non-routine manual. Following Autor et al., (2006), we collapse these five categories to three aggregates: abstract (non-routine analytical and interpersonal), routine (routine cognitive and routine manual), and manual (non-routine manual).

The occupation-level RTI index and its three components for the 13 STEP countries are reported in Appendix Tables A3-A15. Mean RTI values by country and occupation are plotted in Figure 2.1. The first thing that stands out is that in almost all countries, the RTI index is highest for low-paying elementary occupations while it is lowest for high-paying managerial and professional occupations. The low RTI among managerial and professional occupations is in line with what has been observed in the US and most EU countries (Autor and Dorn, 2013; Goos et al., 2014). However, unlike in the US and Europe, where the middle-wage occupations (clerical workers, craft and related trades workers, and machine operators and assemblers) are the most routine-intensive, we find that routine-intensity is highest in low-paying elementary occupations. Although there is some heterogeneity across STEP countries, there is a strong negative correlation between occupation-level RTI and earnings.[9]

[9] The correlation between occupational earnings (standardized within country) and the RTI index, pooling all STEP countries, is -0.60. The relationship is shown in Appendix Figure A1.

## TABLE 2.1
### STEP SURVEY ITEM PER TASK CATEGORY

| Task category | STEP survey item | Variable name | Variable type |
|---|---|---|---|
| **Abstract** (non-routine analytical and interactive) | Thinking at work | m5b_q09 (Wave 1)<br>m5b_q10 (Wave 2)<br>m6b_q10 (Wave 3) | Categorical (1-5) |
| | Learning at work | m5b_q15 (Wave 1)<br>m5b_q17 (Wave 2)<br>m6b_q17 (Wave 3) | Categorical (1-5) |
| | Contact with clients/suppliers | m5b_q04*m5b_q05 (Wave 1)<br>m5b_q05*m5b_q06 (Wave 2)<br>m6b_q05*m6b_q06 (Wave 3) | Categorical (0-10) |
| | Formal presentation to clients | m5b_q10 (Wave 1)<br>m5b_q12 (Wave 2)<br>m6b_q12 (Wave 3) | Binary |
| | Supervising co-workers | m5b_q11 (Wave 1)<br>m5b_q13 (Wave 2)<br>m6b_q13 (Wave 3) | Binary |
| **Routine** (routine cognitive and manual skills) | Routine math tasks | m5a_q18_1—m5a_q18_4 (Wave 1 & 2)<br>m6a_q13_1—m6a_q13_4 (Wave 3) | Categorical (0-4) |
| | Operate | m5b_q08 (Wave 1)<br>m5b_q09 (Wave 2)<br>m6b_q09 (Wave 3) | Binary |
| | Autonomy at work | m5b_q12 (Wave 1)<br>m5b_q14 (Wave 2)<br>m6b_q14 (Wave 3) | Categorical (1-10) |
| | Repetitiveness at work | m5b_q14 (Wave 1)<br>m5b_q16 (Wave 2)<br>m6b_q16 (Wave 3) | Categorical (1-4) |
| **Manual** (non-routine manual skills) | Driving | m5b_q06 (Wave 1)<br>m5b_q07 (Wave 2)<br>m6b_q07 (Wave 3) | Binary |
| | Repair | m5b_q07 (Wave 1)<br>m5b_q08 (Wave 2)<br>m6b_q08 (Wave 3) | Binary |

Source: Authors' elaboration/calculation.
Note: Wave 1 countries are Bolivia, Colombia, Laos, Sri Lanka, Ukraine, Vietnam, and Yunnan province of China. Wave 2 countries are Armenia, Georgia, Ghana, Kenya, and Macedonia. Wave 3 country is the Philippines.

Gasparini et al., (2021) document a similar pattern across six Latin American countries with higher routine intensity in lower paying occupations. If RTI is predictive of a negative employment effect due to RBTC, this pattern suggests RBTC will be associated with declining demand for labor in low-wage occupations, rather than polarization of employment. Indeed, some recent studies find no evidence of polarization in developing countries (e.g., Das and Hilgenstock, 2018; Maloney and Molina, 2016).

## FIGURE 2.1
### ROUTINE TASK INTENSITY INDEX, BY COUNTRY AND OCCUPATION

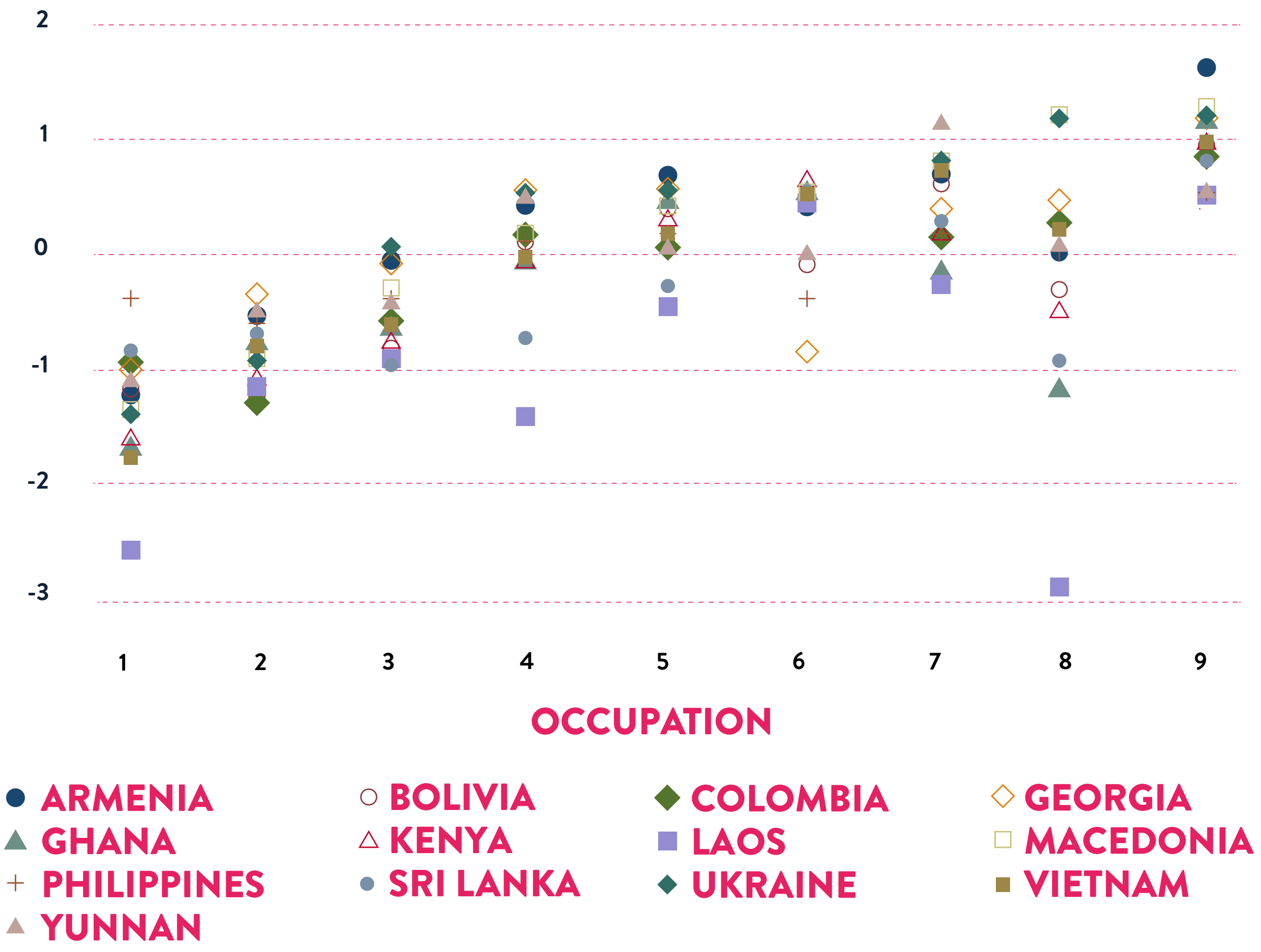

Source: Authors' elaboration/calculation.
Note: Mean Routine Task Intensity index by occupation, for 13 STEP countries. Occupational codes indicate: 1 Managers, 2 Professionals, 3 Technicians and Associate Professionals, 4 Clerical Support Workers, 5 Services and Sales Workers, 6 Skilled Agricultural, Forestry and Fishery Workers, 7 Craft and Related Trades Workers, 8 Plant and Machine Operators and Assemblers, 9 Elementary Occupations.

The data further reveal that within most country-occupation cells, women have a higher RTI than men. Table 2.2 and Figure 2.4 show the average task index measures by country and gender (pooling across occupations). Women have a higher RTI index than men in all countries except the Philippines. Although the Routine task index (column 2 in Table 2.2) is lower for women than for men in every country, the gender difference in the Manual task index (column 4) is much greater and men's high manual task intensity reduces their RTI index. Conversely, the low Manual task index of women's jobs is driving up women's RTI index. The gender gap in the Abstract task index is negative in most countries as well, indicating that women's jobs involve fewer abstract tasks than men's jobs, but it is positive in the three former Soviet Union countries (Armenia, Georgia, and Ukraine) and the Philippines.

## TABLE 2.2
### AVERAGE TASK INTENSITY MEASURES ACROSS ALL WORKERS, BY COUNTRY AND GENDER

| | Obs. | RTI index (1) | Routine task index (2) | Abstract task index (3) | Manual task index (4) |
|---|---|---|---|---|---|
| **Armenia** Male | 373 | -0.37 | 0.13 | -0.03 | 0.53 |
| Female | 626 | 0.25 | -0.08 | 0.02 | -0.35 |
| **Bolivia** Male | 814 | -0.44 | 0.13 | 0.15 | 0.42 |
| Female | 943 | 0.37 | -0.10 | -0.12 | -0.35 |
| **Colombia** Male | 847 | -0.21 | 0.20 | 0.10 | 0.31 |
| Female | 869 | 0.20 | -0.19 | -0.09 | -0.29 |
| **Georgia** Male | 351 | -0.26 | 0.11 | -0.14 | 0.51 |
| Female | 582 | 0.17 | -0.07 | 0.09 | -0.33 |
| **Ghana** Male | 962 | -0.47 | 0.25 | 0.32 | 0.41 |
| Female | 1171 | 0.37 | -0.20 | -0.25 | -0.32 |
| **Kenya** Male | 1339 | -0.12 | 0.09 | 0.06 | 0.15 |
| Female | 1022 | 0.15 | -0.12 | -0.07 | -0.20 |
| **Laos** Male | 918 | -0.18 | 0.15 | 0.15 | 0.19 |
| Female | 1267 | 0.17 | 0.14 | -0.14 | 0.18 |
| **Macedonia** Male | 990 | -0.29 | 0.11 | 0.03 | 0.37 |
| Female | 820 | 0.31 | -0.12 | -0.03 | -0.39 |
| **Philippines** Male | 1007 | 0.01 | 0.03 | -0.04 | 0.06 |
| Female | 681 | -0.01 | -0.04 | 0.05 | -0.09 |
| **Sri Lanka** Male | 912 | -0.24 | 0.06 | 0.05 | 0.25 |
| Female | 647 | 0.35 | -0.09 | -0.07 | -0.37 |
| **Ukraine** Male | 421 | -0.40 | 0.08 | -0.05 | 0.53 |
| Female | 713 | 0.26 | -0.05 | 0.03 | -0.34 |
| **Vietnam** Male | 973 | -0.39 | 0.07 | 0.12 | 0.34 |
| Female | 1359 | 0.29 | -0.05 | -0.19 | -0.25 |
| **Yunnan** Male | 639 | -0.30 | 0.04 | 0.06 | 0.27 |
| Female | 605 | 0.34 | -0.04 | -0.07 | -0.31 |

Source: World Bank STEP household surveys and authors' calculations.
Note: RTI Index = Routine task index – (Abstract task index + Manual task index).

**FIGURE 2.2**
**GENDER GAP IN TASK INTENSITY**

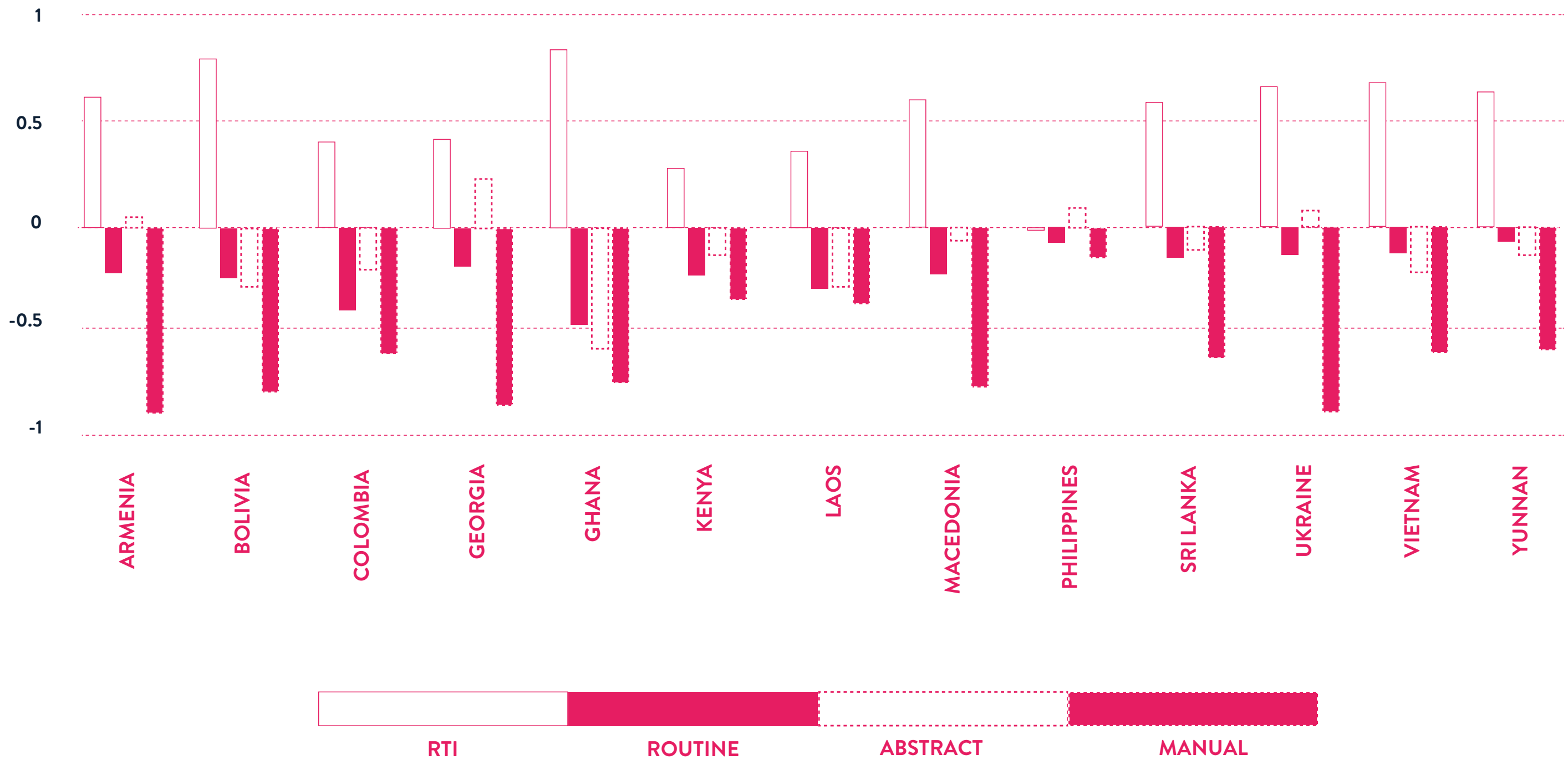

Source: World Bank STEP household surveys and authors' calculations.
Note: Gender gaps measured as female mean index - male mean index. RTI = Routine – (Abstract + Manual).

Women's higher RTI index in the STEP countries is in line with similar patterns across 30 advanced and emerging economies analyzed by Brussevich et al., (2019). Similarly, based on PIAAC data for 24 countries, Brambilla et al., (2021) show that women are less likely to perform abstract tasks (or what they label flexible tasks) than men.

# DECOMPOSITION ANALYSIS

In 12 out of 13 countries included in our analysis, women's routine-task intensity of work exceeds men's, and this also holds within most country-occupation pairs. To assess the role of occupational segregation in accounting for gender differences in routine-task intensity, we start with a simple decomposition analysis. We classify each worker with an RTI score above the own-country median RTI as *high-RTI*. The Gender RTI Gap (GRG) is then defined as the fraction of female *high-RTI* workers minus the fraction of male *high-RTI* workers. We decompose the GRG into a between-occupation and a within-occupation component using the nine 1-digit ISCO-08 occupational groups:

$$GRG = R^f - R^m = \Sigma_j \left\{ \frac{(R_j^f + R_j^m)}{2} \times \left( \frac{F_j}{F} - \frac{M_j}{M} \right) \right\} + \Sigma_j \left\{ \left( \frac{F_j}{F} + \frac{M_j}{M} \right) \times (R_j^f - R_j^m) \right\} \quad (2)$$

In equation (2), $R$ is the share of *high-RTI* jobs in employment, superscripts $f$ and $m$ indicate gender, subscript $j$ indicates occupation, $F$ is the number of female workers, and $M$ is the number of male workers. The first term on the right-hand side captures the between-occupation component and is the sum across occupations of the average share of *high-RTI* jobs within the occupation, multiplied by the gender gap in the occupation's share of employment. The between-occupation term gets larger as women are increasingly overrepresented in occupations with an above-average share of *high-RTI* workers. The second term captures the within-occupation component and is the sum across occupations of each occupation's average share in total female and male employment, multiplied by the gender gap in the within-occupation share of *high-RTI* jobs.

Figure 2.3 summarizes the contribution of the between-occupation and the within-occupation component to the overall gender RTI gap, which ranges from -.01 to .22. Since women's jobs are, on average, more routine task intensive than men's, it is no surprise that the gender RTI gap is positive in all countries, with the exception of the Philippines. In 10 out of 13 countries, the gender gap is almost entirely accounted for by within-occupational gender differences. In Bolivia and Ghana, the between-occupation component accounts for about one third to one half of the total gender gap, while in Kenya it explains almost the entire gap.[10]

## FIGURE 2.3

### DECOMPOSITION OF THE GENDER RTI GAP

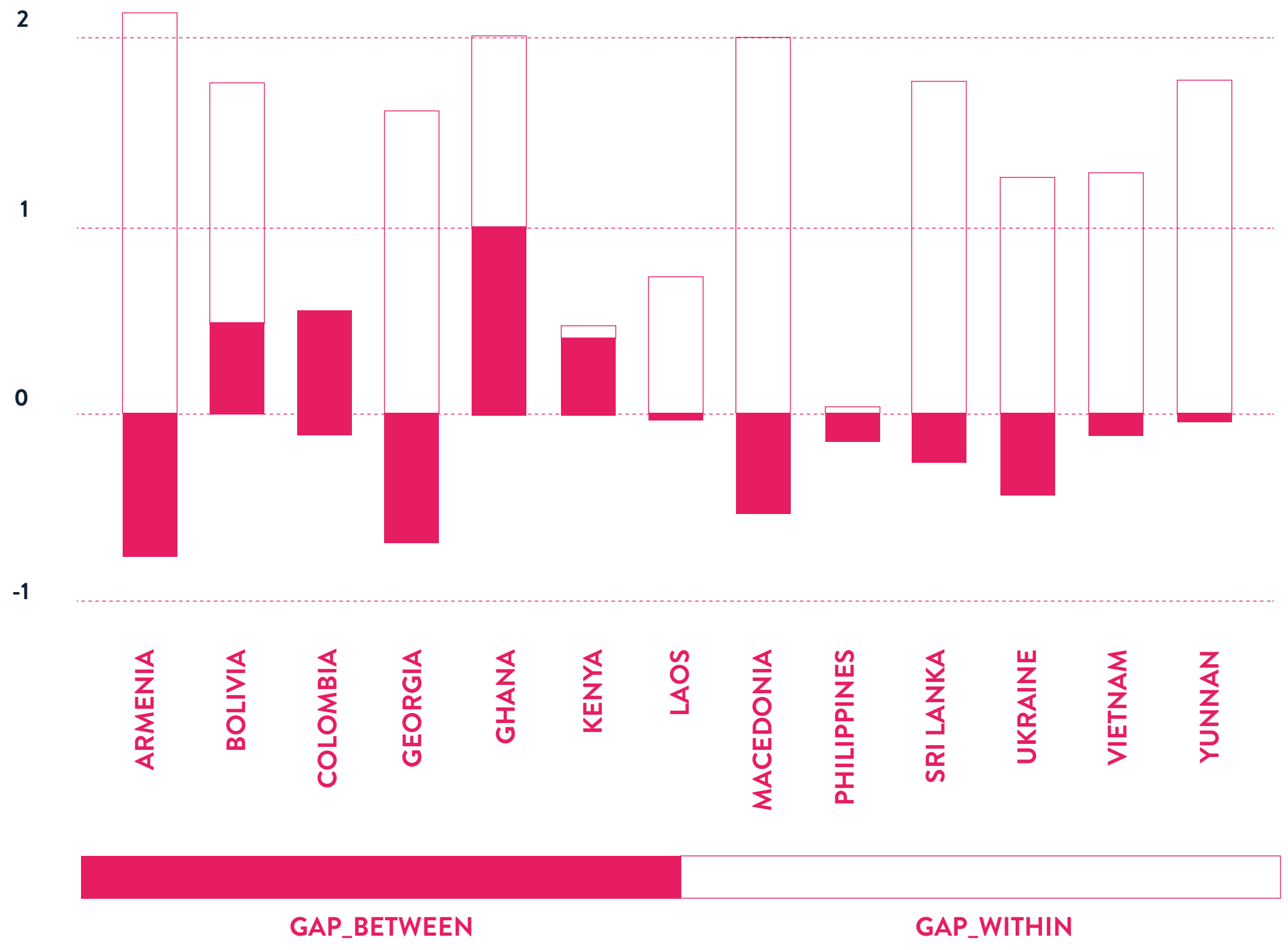

Source: Authors' elaboration/calculation.
Note: Gender RTI gap is the gender gap in the share of workers with an RTI index above the country median RTI. Source: World Bank STEP household surveys and authors' calculations. See equation (2) in the main text.

The fact that the between-occupation contribution is very small or even negative in most countries indicates that occupational segregation cannot explain why women's jobs are more routine task intensive. While we rely on a rather aggregate grouping of occupations, it is still remarkable that differential sorting into these groups explains so little of the gender RTI gap. Further analysis of the data (not reported here) shows that in most of the STEP country labor markets, women are overrepresented among Professionals, Services and Sales Workers, and – to a lesser extent – Clerical Support Workers. While the latter two are somewhat above average in terms of their high-RTI share of workers in most countries, Professionals' RTI is below average, and hence women's overrepresentation in these occupations does not contribute (much) to the overall gender RTI gap. Exceptions are Bolivia, Ghana and Kenya where the between-occupation component is driven by a very high overrepresentation of women among Services and Sales Workers. We further see that in most countries, men are overrepresented among Craft and Related Trades Workers and Plant and Machine Operators and Assemblers, of which the former contains a relatively high share of high-RTI workers. Finally, while Elementary Workers have the highest RTI, women are only slightly overrepresented in this occupation in some of the countries.

[10] In Kenya, the gender RTI gap is driven mainly by the overrepresentation of women among Service and Sales Workers, as well as Elementary Workers, both of which have a relatively high fraction of *high-RTI* jobs.

The gender RTI gap is thus largely driven by women having more routine-intensive jobs then men within the same 1-digit occupational group. The within component is not driven by specific occupations, but rather reflects the fact that women have more routine-intensive jobs within most of the country-occupation pairs. This is further illustrated in Figure 2.4, which shows the relationship between each occupation's share in total employment and the gender RTI gap within the occupation. The gender RTI gap is positive in most country-occupations, including those that account for a large fraction of total employment (such as Services and Sales Workers and Craft and Related Trades Workers) but also most of the smaller occupations.

## FIGURE 2.4
### EMPLOYMENT SHARE AND GENDER RTI GAPS WITHIN OCCUPATION

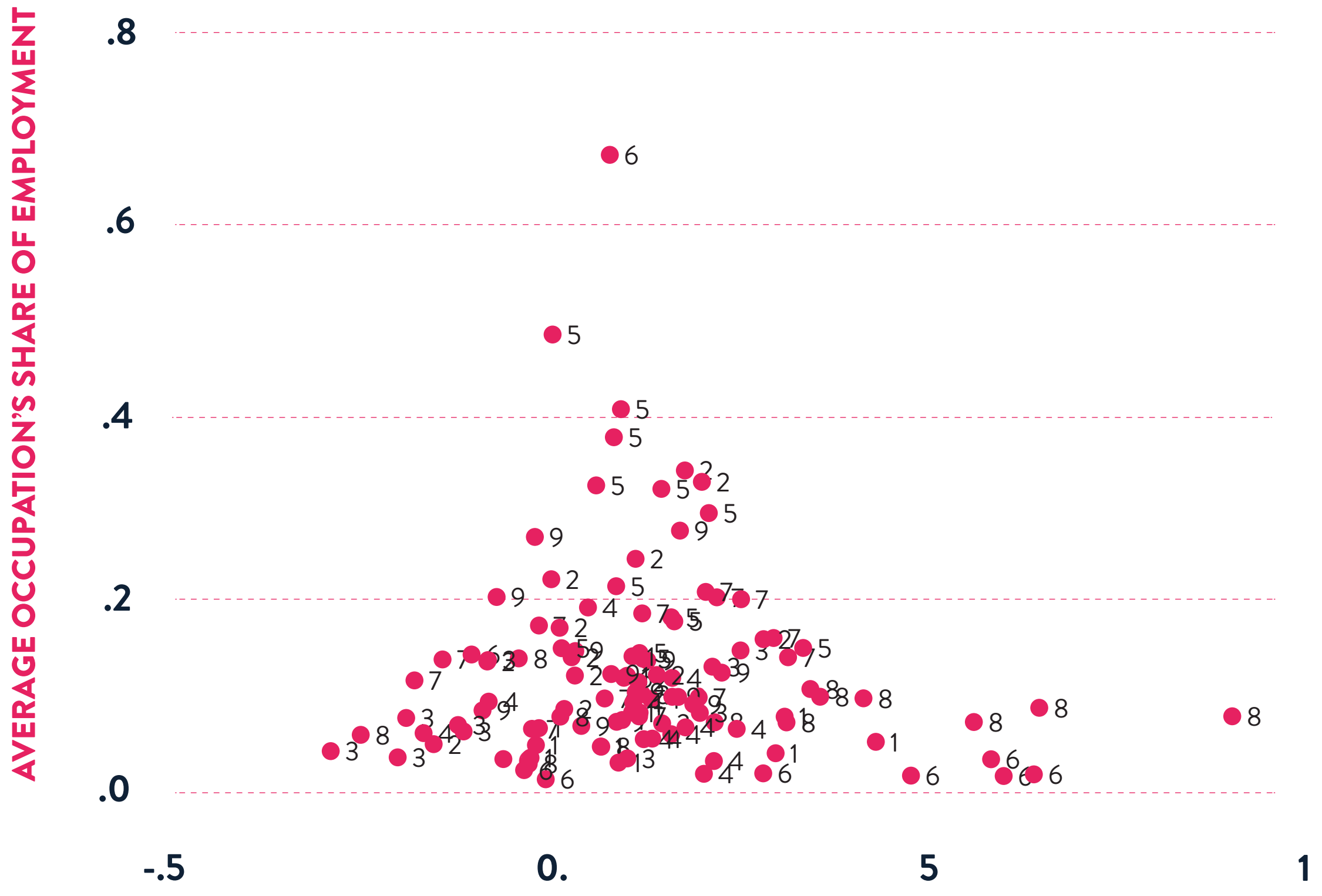

Source: Authors' elaboration/calculation.
Note: Each point represents one country-occupation pair. Labels indicate the 1-digit ISCO group: 1 Managers, 2 Professionals, 3 Technicians and Associate Professionals, 4 Clerical Support Workers, 5 Services and Sales Workers, 6 Skilled Agricultural, Forestry and Fishery Workers, 7 Craft and Related Trades Workers, 8 Plant and Machine Operators and Assemblers, 9 Elementary Occupations.

Figure A2 in the Appendix reports decomposition results based on more detailed, 2-digit occupational groups.[11] The results should be interpreted with caution, since sample sizes in some occupations are very small, but by and large we see that in most countries at least half of the gender RTI gap is still accounted for by within-occupation differences. Colombia is an exception. Here, between-occupation differences explain the entire gender RTI gap.

[11] This analysis excludes the Philippines, for which 2-digit occupation codes are not available.

# CAPITAL, OCCUPATIONAL SORTING, AND THE GENDER GAP IN ROUTINE INTENSITY

To further assess the gender difference in routine-intensity, we regress individuals' RTI index on a *Female* dummy and then add, consecutively, educational attainment (less than high-school, high-school, or more than high-school), work experience (measured as age minus years of education minus six) and its square, ethnicity (an indicator for bilingual or non-native speaker), and occupation. In the regression analysis we use 2-digit occupation dummies. Since 2-digit codes are not included in the data for the Philippines, we exclude this country from the regression analyses.

Estimation results for each country are reported in Appendix Tables A-16 to A-26. Figure 2.5 below summarizes the main findings by plotting the estimated *Female* coefficient for three specifications, by country. Model 1 refers to the specification with no control variables (capturing the unconditional gender gap in the RTI index); in Model 4 we control for education, experience and ethnicity; and in Model 5 we additionally control for 2-digit occupation. Changes in the *Female* coefficient estimate across specifications indicate to what extent human capital variables and occupational sorting account for the unconditional gender gap in the RTI index.

## FIGURE 2.5
### ESTIMATED GENDER GAP IN RTI INDEX ACROSS MODEL SPECIFICATIONS, BY COUNTRY

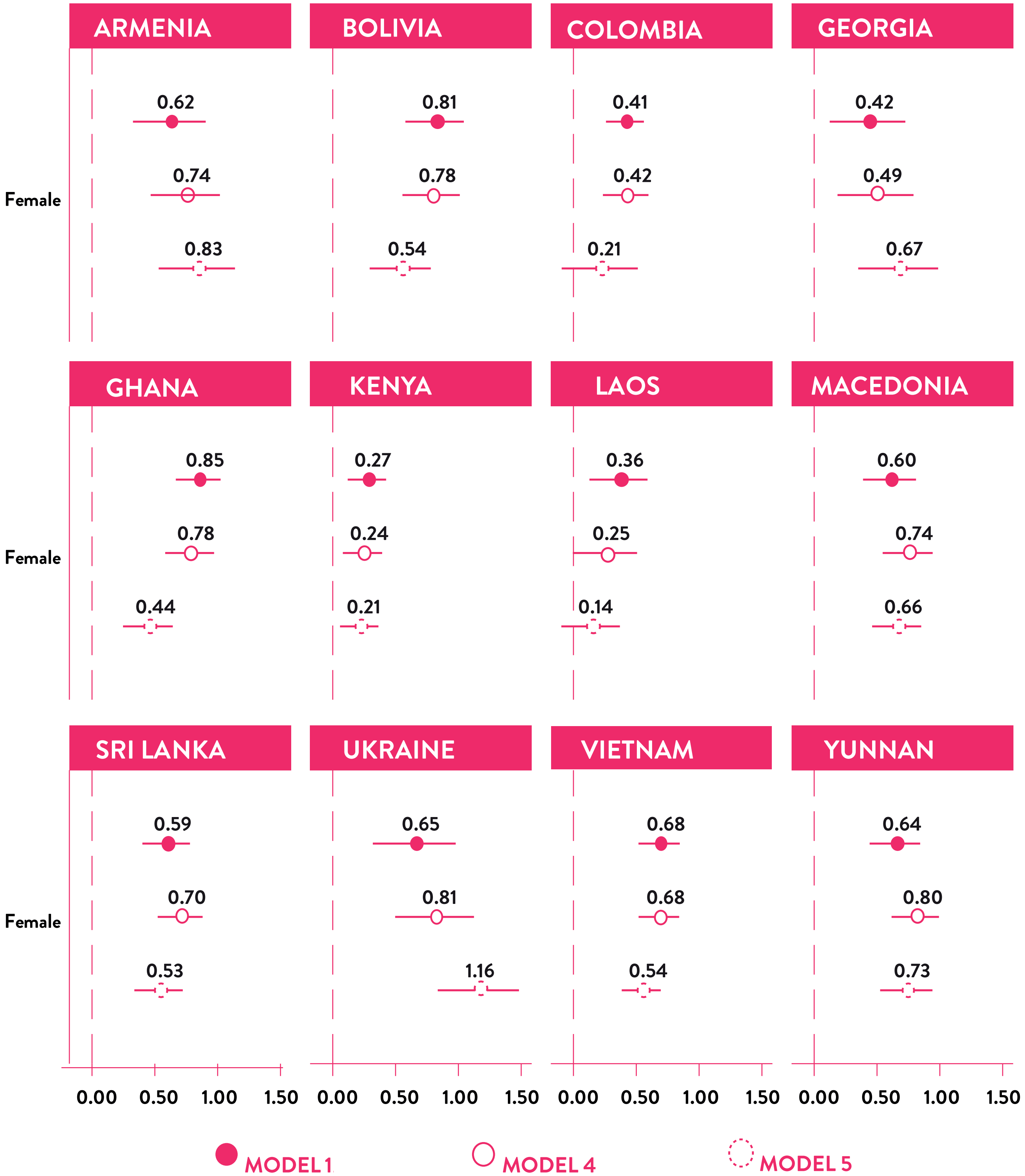

Source: Authors' elaboration/calculation.
Note: Estimated coefficients and 95% confidence interval for Female dummy in OLS regressions where the individual RTI index is the dependent variable. Model 1 refers to the specification with no control variables; in Model 4 we control for education, experience, and ethnic group; in Model 5 we additionally control for 2-digit occupation.

The unconditional gender gap, i.e., the *Female* coefficient estimate in Model 1, is significantly positive in all countries and ranges between .27 in Kenya and .81 in Bolivia, reflecting women's higher routine-task intensity that we also reported in Table 2.2. When we include control variables for education, experience, and ethnicity, the estimated gender gap does not decline substantially – it even increases in 7 out of 12 countries. This indicates that women's higher routine-intensity is not accounted for by gender differences in workers' human capital or ethnicity. Results for Laos are a bit different. Here, inclusion of the same control variables reduces the coefficient estimate for *Female* from .36 to .25.

Controlling for 2-digit occupation (Model 5) reduces the *Female* coefficient in nine of the 12 countries. The effect is most pronounced in Bolivia, Colombia, and Ghana. In Colombia and Laos, the coefficient is no longer statistically significant, indicating that conditional on human capital, ethnic group, and occupational sorting, there is no significant gender difference in RTI. In the other ten countries, women's jobs are significantly more routine-intensive than men's, even conditional on human capital, ethnicity, and 2-digit occupation.

Finally, it is worth noting that in Armenia, Georgia, and Ukraine, however, controlling for occupations leads to an increase in the *Female* coefficient. In these countries, gender differences in occupational sorting have a downward effect on the gender gap in routine-intensity. Within 2-digit occupations, however, women's routine-intensity far exceeds men's.

# CONCLUSIONS

The objective of this chapter was to describe the gender gap in the routine task intensity of jobs in developing country labor markets and to explore the role of occupational segregation and worker characteristics in accounting for the gender gap. Using individual-level harmonized survey data across 13 low- and middle-income countries, we find that women report a higher routine-intensity of their jobs than men. Although men report doing more routine tasks than women, they report even more manual tasks, and this reduces men's relative routine task intensity (RTI).

A decomposition analysis shows that in most countries, the gender RTI gap is largely driven by women doing more routine-intensive work then men *within* the same 1-digit occupational group. This is not driven by specific occupations but reflects the fact that women have more routine-intensive jobs within most country-occupation pairs. Gender differences in occupational choice across 2-digit occupations do account for a part of the gender gap in RTI, but in most countries the contribution is still limited. Differences in human capital and ethnicity also explain little. With the exception of Colombia and Laos, there remains a substantial and statistically significant gender gap in routine task intensity that is unaccounted for by key worker characteristics and occupational choice. These findings are in line with similar evidence for 30 advanced and emerging economies documented by Brussevich et al. (2019).

An important limitation of this study is that we have harmonized data across a limited number of countries, representing only the urban labor markets within those countries, and capturing only one point in time. Nonetheless, we believe that documenting the gender difference in routine-intensity across these low- and middle-income countries contributes to an understanding of gender inequalities in developing countries, in particular related to potential future impacts of new technologies. More research will be needed to assess how automation will affect these labor markets and at what pace the adoption of automation technologies is likely to happen. Since we find that the most routine-intensive occupations are also the lowest paid occupations (in line with other evidence for developing economies documented by Das and Hilgenstock, 2018; Gasparini et al., 2021; Maloney and Molina, 2016), there may be little incentive for employers to invest in the automation of routine tasks. But if they do, this will affect women more than men and will have a disproportionate impact on low-wage workers, which is an important difference with the job polarization documented in the US and Europe.

Our findings imply that aggregate occupation-level measures of occupational task content mask significant gender differences. An important question remains to what extent gender differences in *reporting* of job tasks play a role in the gender routine-intensity differences. While individual level measures of occupational tasks are valuable, they may be less reliable than expert-based measures as used in O*Net. Future work could also explore cross-country differences in income, sectoral structure, and female labor force participation, as well as employer's gender biases, to learn more about the nature of gender task segregation in low- and middle-income countries.

# APPENDIX

## FIGURE A1

EARNINGS AND ROUTINE INTENSITY
BY COUNTRY-OCCUPATION PAIRS

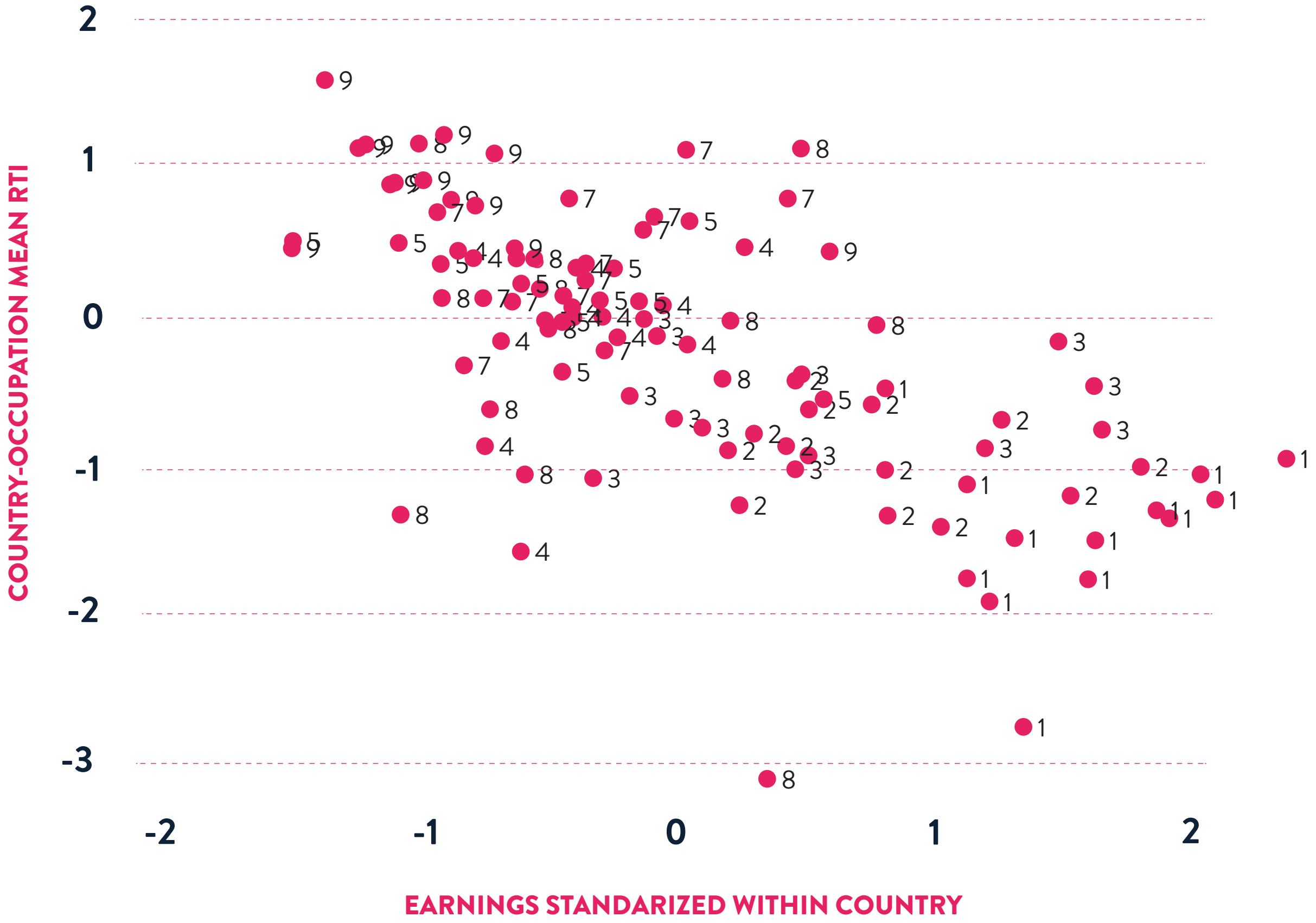

Source: Authors' elaboration/calculation.
Note: Estimated coefficients and 95% confidence interval for Female dummy in OLS regressions where the individual RTI index is the dependent variable.
Model 1 refers to the specification with no control variables; in Model 4 we control for education, experience, and ethnic group; in Model 5 we additionally control for 2-digit occupation.

**FIGURE A2**

**DECOMPOSITION OF THE GENDER RTI GAP BASED ON 2-DIGIT OCCUPATIONS**

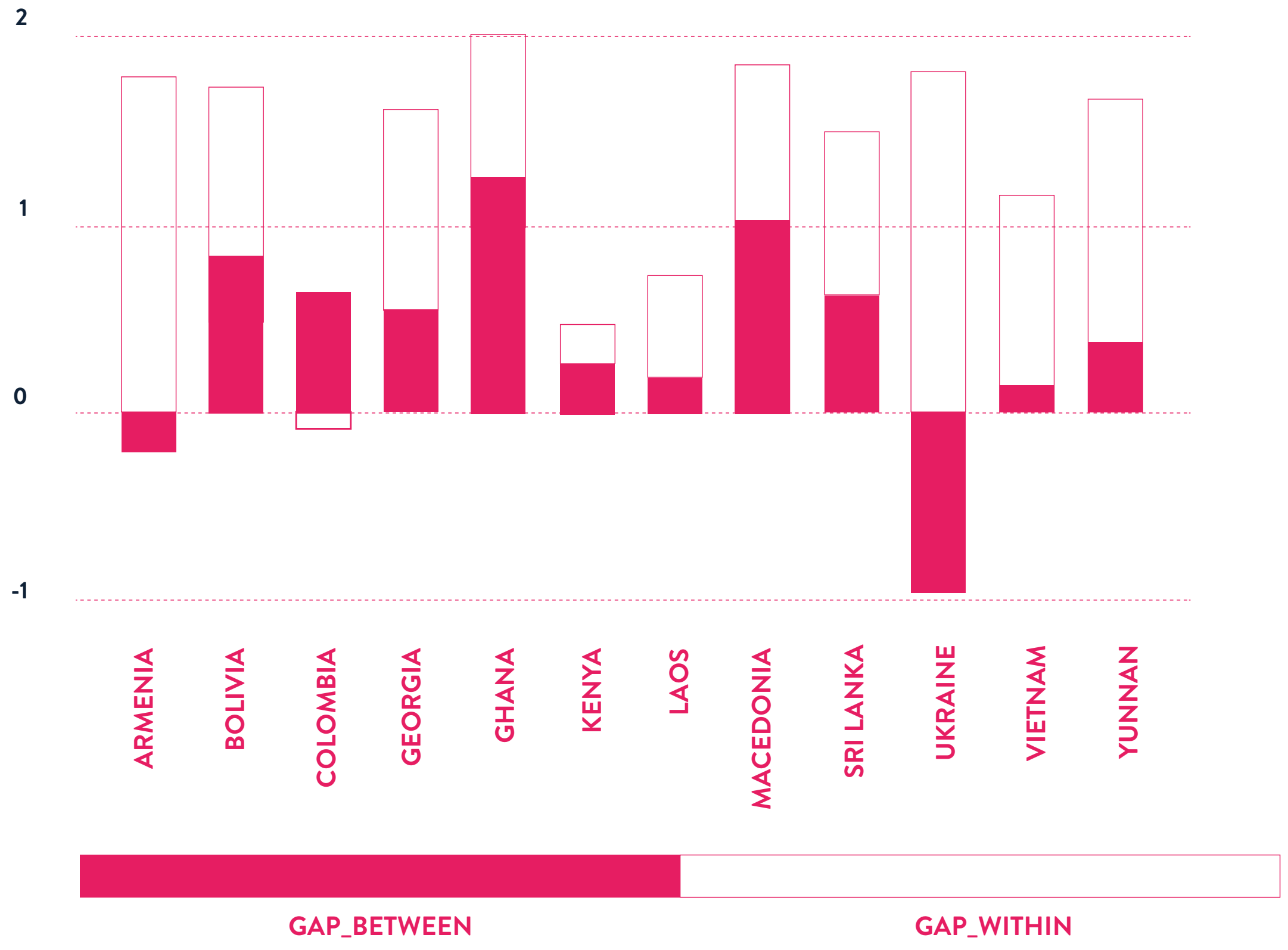

Source: Authors' elaboration/calculation.
Note: Gender RTI gap is the gender gap in the share of workers with an RTI index above the country median RTI. Source: World Bank STEP household surveys and authors' calculations. See equation (2) in the main text.

## TABLE A1
### ARMENIA-TASK MEASURES OF MAJOR OCCUPATION GROUPS BY GENDER

| | RTI index | Routine task index | Abstract task index | Manual task index |
|---|---|---|---|---|
| **MANAGERS** All | -1.34 | -0.09 | 0.77 | 0.48 |
| Male | -1.97 | -0.05 | 0.92 | 0.99 |
| Female | -0.37 | -0.15 | 0.53 | -0.30 |
| **PROFESSIONALS** All | -0.61 | -0.16 | 0.49 | -0.04 |
| Male | -1.42 | -0.10 | 0.58 | 0.73 |
| Female | -0.35 | -0.17 | 0.46 | -0.29 |
| **TECHNICIANS AND ASSOCIATE PROFESSIONALS** All | -0.13 | 0.03 | 0.20 | -0.04 |
| Male | -0.45 | 0.04 | 0.13 | 0.36 |
| Female | 0.13 | 0.02 | 0.25 | -0.36 |
| **CLERICAL SUPPORT WORKERS** All | 0.33 | 0.15 | 0.03 | -0.21 |
| Male | -0.78 | 0.02 | 0.29 | 0.51 |
| Female | 0.58 | 0.19 | -0.03 | -0.36 |
| **SERVICE AND SALES WORKERS** | | | | |
| All | 0.63 | 0.01 | -0.40 | -0.25 |
| Male | 0.38 | -0.05 | -0.43 | 0.01 |
| Female | 0.81 | 0.01 | -0.38 | -0.42 |
| **SKILLED AGRICULTURAL, FORESTRY AND FISHERY WORKERS** | | | | |
| All | 0.35 | -0.76 | -1.08 | -0.03 |
| Male | -1.01 | -1.13 | -0.57 | 0.45 |
| Female | 2.19 | -0.25 | -1.76 | -0.68 |
| **CRAFT AND RELATED TRADES WORKERS** | | | | |
| All | 0.67 | 0.38 | -0.52 | 0.23 |
| Male | 0.51 | 0.44 | -0.43 | 0.36 |
| Female | 1.44 | 0.06 | -0.98 | -0.40 |
| **PLANT AND MACHINE OPERATORS AND ASSEMBLERS** | | | | |
| All | -0.05 | 0.45 | -0.53 | 1.03 |
| Male | -0.14 | 0.47 | -0.51 | 1.12 |
| Female | 1.80 | 0.11 | -1.01 | -0.68 |
| **ELEMENTARY OCCUPATIONS** | | | | |
| All | 1.58 | -0.10 | -1.28 | -0.41 |
| Male | 1.92 | 0.57 | -1.08 | 0.28 |
| Female | 1.47 | -0.32 | -1.34 | -0.45 |

Source: Authors' elaboration/calculation.

## TABLE A2

### BOLIVIA-TASK MEASURES OF MAJOR OCCUPATION GROUPS BY GENDER

| | RTI index | Routine task index | Abstract task index | Manual task index |
|---|---|---|---|---|
| **MANAGERS** All | -1.28 | 0.18 | 1.22 | 0.24 |
| Male | -1.71 | 0.22 | 1.39 | 0.54 |
| Female | -0.46 | 0.09 | 0.90 | -0.35 |
| **PROFESSIONALS** All | -1.33 | -0.24 | 1.04 | 0.06 |
| Male | -1.59 | -0.17 | 0.93 | 0.49 |
| Female | -1.15 | -0.30 | 1.12 | -0.27 |
| **TECHNICIANS AND ASSOCIATE PROFESSIONALS** All | -0.92 | 0.09 | 0.56 | 0.45 |
| Male | -1.12 | 0.12 | 0.42 | 0.82 |
| Female | -0.62 | 0.05 | 0.77 | -0.10 |
| **CLERICAL SUPPORT WORKERS** All | 0.01 | 0.24 | 0.45 | -0.22 |
| Male | -0.16 | 0.25 | 0.54 | -0.12 |
| Female | 0.18 | 0.22 | 0.37 | -0.32 |
| **SERVICE AND SALES WORKERS** All | 0.33 | -0.16 | -0.25 | -0.24 |
| Male | -0.27 | -0.17 | -0.07 | 0.17 |
| Female | 0.53 | -0.15 | -0.31 | -0.38 |
| **SKILLED AGRICULTURAL, FORESTRY AND FISHERY WORKERS** All | -0.17 | -0.37 | -0.19 | -0.01 |
| Male | -0.53 | -0.30 | 0.02 | 0.22 |
| Female | 0.52 | -0.50 | -0.58 | -0.44 |
| **CRAFT AND RELATED TRADES WORKERS** All | 0.59 | 0.26 | -0.25 | -0.08 |
| Male | 0.19 | 0.33 | -0.04 | 0.18 |
| Female | 1.13 | 0.16 | -0.53 | -0.43 |
| **PLANT AND MACHINE OPERATORS AND ASSEMBLERS** All | -0.41 | 0.41 | -0.45 | 1.27 |
| Male | -0.53 | 0.42 | -0.47 | 1.42 |
| Female | 0.47 | 0.33 | -0.36 | 0.22 |
| **ELEMENTARY OCCUPATIONS** All | 0.89 | -0.19 | -0.84 | -0.23 |
| Male | 0.68 | 0.01 | -0.70 | 0.04 |
| Female | 1.02 | -0.31 | -0.93 | -0.40 |

Source: Authors' elaboration/calculation.

## TABLE A3
### COLOMBIA-TASK MEASURES OF MAJOR OCCUPATION GROUPS BY GENDER

| | RTI index | Routine task index | Abstract task index | Manual task index |
|---|---|---|---|---|
| **MANAGERS** | | | | |
| All | -1.04 | 0.18 | 0.70 | 0.53 |
| Male | -1.12 | 0.26 | 0.68 | 0.70 |
| Female | -0.96 | 0.09 | 0.72 | 0.33 |
| **PROFESSIONALS** | | | | |
| All | -1.39 | -0.49 | 1.13 | -0.23 |
| Male | -1.56 | -0.41 | 1.23 | -0.09 |
| Female | -1.27 | -0.55 | 1.05 | -0.33 |
| **TECHNICIANS AND ASSOCIATE PROFESSIONALS** | | | | |
| All | -0.67 | 0.19 | 0.66 | 0.19 |
| Male | -0.78 | 0.34 | 0.68 | 0.44 |
| Female | -0.42 | -0.16 | 0.63 | -0.37 |
| **CLERICAL SUPPORT WORKERS** | | | | |
| All | 0.07 | 0.26 | 0.30 | -0.11 |
| Male | 0.14 | 0.34 | 0.11 | 0.09 |
| Female | 0.01 | 0.20 | 0.44 | -0.25 |
| **SERVICE AND SALES WORKERS** | | | | |
| All | -0.02 | -0.17 | -0.06 | -0.09 |
| Male | -0.63 | -0.13 | 0.13 | 0.37 |
| Female | 0.28 | -0.19 | -0.16 | -0.31 |
| **CRAFT AND RELATED TRADES WORKERS** | | | | |
| All | 0.11 | 0.25 | 0.10 | 0.05 |
| Male | 0.09 | 0.57 | 0.21 | 0.27 |
| Female | 0.13 | -0.19 | -0.06 | -0.26 |
| **PLANT AND MACHINE OPERATORS AND ASSEMBLERS** | | | | |
| All | 0.19 | 0.56 | -0.36 | 0.73 |
| Male | -0.21 | 0.54 | -0.26 | 1.01 |
| Female | 1.73 | 0.63 | -0.75 | -0.35 |
| **ELEMENTARY OCCUPATIONS** | | | | |
| All | 0.79 | -0.20 | -0.74 | -0.25 |
| Male | 0.80 | 0.03 | -0.62 | -0.16 |
| Female | 0.78 | -0.39 | -0.84 | -0.33 |

Source: Authors' elaboration/calculation.

## TABLE A4
### GEORGIA-TASK MEASURES OF MAJOR OCCUPATION GROUPS BY GENDER

| | RTI index | Routine task index | Abstract task index | Manual task index |
|---|---|---|---|---|
| **MANAGERS** All | -1.11 | 0.03 | 0.68 | 0.46 |
| Male | -1.31 | 0.17 | 0.62 | 0.86 |
| Female | -0.88 | -0.13 | 0.75 | 0.01 |
| **PROFESSIONALS** All | -0.42 | -0.21 | 0.44 | -0.23 |
| Male | -1.30 | -0.31 | 0.55 | 0.43 |
| Female | -0.19 | -0.18 | 0.41 | -0.39 |
| **TECHNICIANS AND ASSOCIATE PROFESSIONALS** All | -0.16 | 0.35 | 0.30 | 0.20 |
| Male | -0.46 | 0.38 | 0.21 | 0.63 |
| Female | 0.11 | 0.31 | 0.38 | -0.18 |
| **CLERICAL SUPPORT WORKERS** All | 0.47 | 0.02 | -0.20 | -0.26 |
| Male | 0.21 | 0.52 | -0.29 | 0.60 |
| Female | 0.52 | -0.08 | -0.18 | -0.42 |
| **SERVICE AND SALES WORKERS** | | | | |
| All | 0.51 | 0.01 | -0.31 | -0.24 |
| Male | 0.30 | -0.10 | -0.44 | 0.04 |
| Female | 0.64 | 0.08 | -0.21 | -0.32 |
| **SKILLED AGRICULTURAL, FORESTRY AND FISHERY WORKERS** | | | | |
| All | -0.95 | -0.15 | -0.29 | 1.09 |
| Male | -1.12 | -0.09 | -0.19 | 1.22 |
| Female | 1.12 | -0.88 | -1.56 | -0.44 |
| **CRAFT AND RELATED TRADES WORKERS** | | | | |
| All | 0.36 | 0.30 | -0.56 | 0.50 |
| Male | 0.31 | 0.34 | -0.59 | 0.62 |
| Female | 064 | 0.14 | -0.43 | -0.06 |
| **PLANT AND MACHINE OPERATORS AND ASSEMBLERS** | | | | |
| All | 0.40 | 0.41 | -0.73 | 0.74 |
| Male | 0.18 | 0.30 | -0.71 | 0.83 |
| Female | 3.12 | 1.74 | -0.93 | -0.44 |
| **ELEMENTARY OCCUPATIONS** | | | | |
| All | 1.13 | -0.07 | -1.04 | -0.16 |
| Male | 0.96 | 0.28 | -0.92 | 0.24 |
| Female | 1.24 | -0.29 | -1.12 | -0.41 |

Source: Authors' elaboration/calculation.

## TABLE A5
### GHANA—TASK MEASURES OF MAJOR OCCUPATION GROUPS BY GENDER

| | RTI index | Routine task index | Abstract task index | Manual task index |
|---|---|---|---|---|
| **MANAGERS** | | | | |
| All | -1.82 | 0.44 | 1.22 | 1.04 |
| Male | -2.06 | 0.60 | 1.45 | 1.20 |
| Female | -0.87 | -0.20 | 0.30 | 0.37 |
| **PROFESSIONALS** | | | | |
| All | -0.86 | 0.28 | 1.16 | -0.02 |
| Male | -1.04 | 0.24 | 1.16 | 0.12 |
| Female | -0.53 | 0.36 | 1.16 | -0.27 |
| **TECHNICIANS AND ASSOCIATE PROFESSIONALS** | | | | |
| All | -0.75 | 0.43 | 0.83 | 0.35 |
| Male | -0.67 | 0.64 | 0.75 | 0.56 |
| Female | -1.02 | -0.28 | 1.11 | -0.37 |
| **CLERICAL SUPPORT WORKERS** | | | | |
| All | -0.18 | 0.38 | 0.62 | -0.07 |
| Male | -0.62 | 0.28 | 0.73 | 0.17 |
| Female | 0.24 | 0.46 | 0.52 | -0.29 |
| **SERVICE AND SALES WORKERS** | | | | |
| All | 0.39 | -0.22 | -0.36 | -0.25 |
| Male | 0.00 | 0.08 | -0.07 | 0.15 |
| Female | 0.49 | -0.29 | -0.43 | -0.35 |
| **SKILLED AGRICULTURAL, FORESTRY AND FISHERY WORKERS** | | | | |
| All | 0.47 | -0.42 | -0.60 | -0.29 |
| Male | 0.42 | -0.35 | -0.53 | -0.24 |
| Female | 0.54 | -0.53 | -0.70 | -0.37 |
| **CRAFT AND RELATED TRADES WORKERS** | | | | |
| All | -0.21 | 0.17 | 0.19 | 0.19 |
| Male | -0.79 | 0.40 | 0.51 | 0.68 |
| Female | 0.33 | -0.05 | -0.11 | -0.27 |
| **PLANT AND MACHINE OPERATORS AND ASSEMBLERS** | | | | |
| All | -1.31 | 0.52 | 0.07 | 1.76 |
| Male | -1.31 | 0.52 | 0.07 | 1.76 |
| Female | - | - | - | - |
| **ELEMENTARY OCCUPATIONS** | | | | |
| All | 1.09 | 0.23 | -0.67 | -0.19 |
| Male | 1.13 | 0.55 | -0.50 | -0.08 |
| Female | 1.05 | -0.22 | -0.91 | -0.35 |

Source: Authors' elaboration/calculation.

**TABLE A6**

**KENYA—TASK MEASURES OF MAJOR OCCUPATION GROUPS BY GENDER**

| | RTI index | Routine task index | Abstract task index | Manual task index |
|---|---|---|---|---|
| **MANAGERS** | | | | |
| All | -1.74 | 0.01 | 1.13 | 0.61 |
| Male | -1.73 | 0.10 | 1.29 | 0.54 |
| Female | -1.75 | -0.10 | 0.96 | 0.69 |
| **PROFESSIONALS** | | | | |
| All | -1.19 | 0.11 | 1.02 | 0.27 |
| Male | -1.48 | 0.04 | 1.03 | 0.48 |
| Female | -0.67 | 0.23 | 1.00 | -0.10 |
| **TECHNICIANS AND ASSOCIATE PROFESSIONALS** | | | | |
| All | -0.87 | 0.19 | 0.69 | 0.37 |
| Male | -0.81 | 0.25 | 0.65 | 0.41 |
| Female | -1.02 | 0.03 | 0.78 | 0.27 |
| **CLERICAL SUPPORT WORKERS** | | | | |
| All | -0.18 | 0.03 | 0.14 | 0.06 |
| Male | -0.44 | 0.09 | 0.32 | 0.21 |
| Female | 0.13 | -0.05 | -0.07 | -0.11 |
| **SERVICE AND SALES WORKERS** | | | | |
| All | 0.23 | -0.13 | -0.13 | -0.23 |
| Male | 0.18 | -0.12 | -0.16 | -0.13 |
| Female | 0.27 | -0.15 | -0.10 | -0.32 |
| **SKILLED AGRICULTURAL, FORESTRY AND FISHERY WORKERS** | | | | |
| All | 0.58 | -0.38 | -0.84 | -0.12 |
| Male | 0.58 | -0.19 | -0.73 | -0.04 |
| Female | 0.58 | -0.75 | -1.05 | -0.28 |
| **CRAFT AND RELATED TRADES WORKERS** | | | | |
| All | 0.13 | 0.46 | 0.13 | 0.20 |
| Male | 0.25 | 0.61 | 0.09 | 0.27 |
| Female | -0.24 | -0.02 | 0.27 | -0.04 |
| **PLANT AND MACHINE OPERATORS AND ASSEMBLERS** | | | | |
| All | -0.61 | 0.58 | -0.14 | 1.33 |
| Male | -0.64 | 0.60 | -0.09 | 1.33 |
| Female | -0.48 | 0.47 | -0.40 | 1.34 |
| **ELEMENTARY OCCUPATIONS** | | | | |
| All | 0.90 | -0.21 | -0.85 | -0.26 |
| Male | 0.84 | -0.12 | -0.77 | -0.19 |
| Female | 0.96 | -0.30 | -0.93 | -0.33 |

Source: Authors' elaboration/calculation.

## TABLE A7
### LAOS—TASK MEASURES OF MAJOR OCCUPATION GROUPS BY GENDER

| | RTI index | Routine task index | Abstract task index | Manual task index |
|---|---|---|---|---|
| **MANAGERS** All | -2.73 | 0.11 | 1.80 | 1.04 |
| Male | -3.41 | 0.12 | 2.06 | 1.48 |
| Female | -0.85 | 0.08 | 1.09 | -0.16 |
| **PROFESSIONALS** All | -1.25 | 0.37 | 1.59 | 0.04 |
| Male | -1.19 | 0.50 | 1.55 | 0.15 |
| Female | -1.32 | 0.21 | 1.64 | -0.11 |
| **TECHNICIANS AND ASSOCIATE PROFESSIONALS** All | -1.01 | 0.48 | 1.31 | 0.19 |
| Male | -1.53 | 0.55 | 1.55 | 0.53 |
| Female | -0.44 | 0.42 | 1.04 | -0.19 |
| **CLERICAL SUPPORT WORKERS** All | -1.56 | 0.38 | 1.45 | 0.49 |
| Male | -2.44 | -0.05 | 1.72 | 0.68 |
| Female | -0.92 | 0.70 | 1.26 | 0.35 |
| **SERVICE AND SALES WORKERS** | | | | |
| All | -0.54 | 0.03 | 0.46 | 0.12 |
| Male | -1.33 | 0.17 | 0.82 | 0.67 |
| Female | -0.25 | -0.02 | 0.32 | -0.09 |
| **SKILLED AGRICULTURAL, FORESTRY AND FISHERY WORKERS** | | | | |
| All | 0.39 | -0.10 | -0.31 | -0.17 |
| Male | 0.35 | 0.04 | -0.19 | -0.12 |
| Female | 0.42 | -0.24 | -0.43 | -0.23 |
| **CRAFT AND RELATED TRADES WORKERS** | | | | |
| All | -0.32 | 0.36 | 0.17 | 0.50 |
| Male | -0.65 | 0.80 | 0.41 | 1.04 |
| Female | -0.01 | -0.05 | -0.05 | 0.00 |
| **PLANT AND MACHINE OPERATORS AND ASSEMBLERS** | | | | |
| All | -3.07 | -0.06 | 0.23 | 2.79 |
| Male | -3.55 | 0.14 | 0.27 | 3.42 |
| Female | -0.79 | -1.01 | 0.02 | -0.24 |
| **ELEMENTARY OCCUPATIONS** | | | | |
| All | 0.44 | 0.22 | -0.10 | -0.13 |
| Male | 0.28 | 0.33 | 0.13 | -0.08 |
| Female | 0.71 | 0.05 | -0.46 | -0.20 |

Source: Authors' elaboration/calculation.

## TABLE A8
### MACEDONIA—TASK MEASURES OF MAJOR OCCUPATION GROUPS BY GENDER

| | RTI index | Routine task index | Abstract task index | Manual task index |
|---|---|---|---|---|
| **MANAGERS** All | -1.47 | -0.26 | 0.70 | 0.51 |
| Male | -1.76 | -0.17 | 0.75 | 0.83 |
| Female | -1.05 | -0.38 | 0.62 | 0.05 |
| **PROFESSIONALS** All | -1.00 | -0.32 | 0.69 | -0.01 |
| Male | -1.44 | -0.30 | 0.77 | 0.36 |
| Female | -0.69 | -0.33 | 0.64 | -0.27 |
| **TECHNICIANS AND ASSOCIATE PROFESSIONALS** All | -0.38 | 0.11 | 0.30 | 0.18 |
| Male | -0.92 | 0.23 | 0.42 | 0.73 |
| Female | 0.08 | 0.00 | 0.20 | -0.28 |
| **CLERICAL SUPPORT WORKERS** All | 0.08 | -0.07 | 0.05 | -0.20 |
| Male | -0.15 | 0.12 | 0.05 | 0.22 |
| Female | 0.25 | -0.20 | 0.05 | -0.50 |
| **SERVICE AND SALES WORKERS** | | | | |
| All | 0.36 | -0.01 | -0.24 | -0.13 |
| Male | 0.07 | 0.00 | -0.28 | 0.21 |
| Female | 0.65 | -0.02 | -0.19 | -0.48 |
| **SKILLED AGRICULTURAL, FORESTRY AND FISHERY WORKERS** | | | | |
| All | 0.48 | -0.25 | -0.95 | 0.22 |
| Male | 0.04 | -0.12 | -0.75 | 0.58 |
| Female | 1.59 | -0.55 | -1.45 | -0.69 |
| **CRAFT AND RELATED TRADES WORKERS** | | | | |
| All | 0.79 | 0.40 | -0.41 | 0.02 |
| Male | 0.46 | 0.41 | -0.26 | 0.21 |
| Female | 1.90 | 0.35 | -0.93 | -0.62 |
| **PLANT AND MACHINE OPERATORS AND ASSEMBLERS** | | | | |
| All | 1.16 | 0.34 | -0.77 | -0.06 |
| Male | 0.57 | 0.37 | -0.55 | 0.36 |
| Female | 2.00 | 0.28 | -1.08 | -0.65 |
| **ELEMENTARY OCCUPATIONS** | | | | |
| All | 1.23 | 0.02 | -0.97 | -0.25 |
| Male | 0.94 | 0.28 | -0.78 | 0.12 |
| Female | 1.51 | -0.22 | -1.14 | -0.58 |

Source: Authors' elaboration/calculation.

## TABLE A9
### PHILIPPINES—TASK MEASURES OF MAJOR OCCUPATION GROUPS BY GENDER

| | RTI index | Routine task index | Abstract task index | Manual task index |
|---|---|---|---|---|
| **MANAGERS** | | | | |
| All | -0.47 | -0.25 | 0.61 | -0.40 |
| Male | -0.61 | -0.32 | 0.64 | -0.35 |
| Female | -0.31 | -0.18 | 0.58 | -0.45 |
| **PROFESSIONALS** | | | | |
| All | -0.68 | -0.16 | 0.78 | -0.26 |
| Male | -0.62 | -0.12 | 0.75 | -0.26 |
| Female | -0.77 | -0.23 | 0.82 | -0.28 |
| **TECHNICIANS AND ASSOCIATE PROFESSIONALS** | | | | |
| All | -0.47 | -0.23 | 0.50 | -0.26 |
| Male | -0.37 | -0.30 | 0.32 | -0.25 |
| Female | -0.61 | -0.13 | 0.75 | -0.27 |
| **CLERICAL SUPPORT WORKERS** | | | | |
| All | 0.01 | 0.04 | 0.02 | 0.00 |
| Male | -0.21 | 0.04 | 0.14 | 0.11 |
| Female | 0.33 | 0.03 | -0.14 | -0.15 |
| **SERVICE AND SALES WORKERS** All | 0.11 | 0.09 | -0.16 | 0.15 |
| Male | 0.09 | 0.10 | -0.24 | 0.26 |
| Female | 0.14 | 0.08 | -0.05 | -0.01 |
| **SKILLED AGRICULTURAL, FORESTRY AND FISHERY WORKERS** All | -0.47 | 0.31 | 0.67 | 0.11 |
| Male | -0.47 | 0.31 | 0.67 | 0.11 |
| Female | - | - | - | - |
| **CRAFT AND RELATED TRADES WORKERS** All | 0.14 | -0.11 | -0.24 | -0.01 |
| Male | 0.34 | 0.04 | -0.38 | 0.08 |
| Female | -0.15 | -0.32 | -0.03 | -0.13 |
| **PLANT AND MACHINE OPERATORS AND ASSEMBLERS** All | -0.07 | -0.18 | -0.22 | 0.11 |
| Male | -0.12 | -0.16 | -0.23 | 0.20 |
| Female | 0.00 | -0.21 | -0.20 | -0.01 |
| **ELEMENTARY OCCUPATIONS** All | 0.46 | 0.27 | -0.35 | 0.16 |
| Male | 0.48 | 0.30 | -0.36 | 0.17 |
| Female | 0.44 | 0.23 | -0.34 | 0.13 |

Source: Authors' elaboration/calculation.

## TABLE A10

### SRI LANKA—TASK MEASURES OF MAJOR OCCUPATION GROUPS BY GENDER

| | RTI index | Routine task index | Abstract task index | Manual task index |
|---|---|---|---|---|
| **MANAGERS** | | | | |
| All | -0.94 | 0.04 | 0.54 | 0.44 |
| Male | -1.38 | 0.13 | 0.75 | 0.76 |
| Female | 0.27 | -0.21 | -0.05 | -0.42 |
| **PROFESSIONALS** | | | | |
| All | -0.77 | 0.20 | 1.06 | -0.08 |
| Male | -1.25 | 0.15 | 1.13 | 0.26 |
| Female | -0.54 | 0.23 | 1.02 | -0.25 |
| **TECHNICIANS AND ASSOCIATE PROFESSIONALS** | | | | |
| All | -1.07 | -0.24 | 0.68 | 0.15 |
| Male | -0.99 | -0.37 | 0.69 | -0.07 |
| Female | -1.22 | 0.01 | 0.67 | 0.57 |
| **CLERICAL SUPPORT WORKERS** | | | | |
| All | -0.86 | 0.15 | 0.67 | 0.33 |
| Male | -1.05 | 0.18 | 0.73 | 0.49 |
| Female | -0.63 | 0.12 | 0.60 | 0.14 |
| **SERVICE AND SALES WORKERS** All | -0.36 | -0.04 | 0.20 | 0.12 |
| Male | -0.50 | 0.01 | 0.17 | 0.34 |
| Female | -0.14 | -0.10 | 0.24 | -0.20 |
| **SKILLED AGRICULTURAL, FORESTRY AND FISHERY WORKERS** All | 0.49 | -0.39 | -0.45 | -0.43 |
| Male | 0.50 | -0.28 | -0.47 | -0.31 |
| Female | 0.48 | -0.51 | -0.44 | -0.56 |
| **CRAFT AND RELATED TRADES WORKERS** All | 0.25 | 0.32 | 0.05 | 0.01 |
| Male | 0.00 | 0.34 | 0.14 | 0.19 |
| Female | 0.79 | 0.27 | -0.14 | -0.39 |
| **PLANT AND MACHINE OPERATORS AND ASSEMBLERS** All | -1.04 | 0.12 | -0.09 | 1.25 |
| Male | -1.40 | 0.07 | 0.00 | 1.46 |
| Female | 2.02 | 0.55 | -0.89 | -0.57 |
| **ELEMENTARY OCCUPATIONS** All | 0.75 | -0.16 | -0.56 | -0.34 |
| Male | 0.58 | -0.04 | -0.44 | -0.18 |
| Female | 0.96 | -0.31 | -0.72 | -0.55 |

Source: Authors' elaboration/calculation.

## TABLE A11

### UKRAINE—TASK MEASURES OF MAJOR OCCUPATION GROUPS BY GENDER

| | RTI index | Routine task index | Abstract task index | Manual task index |
|---|---|---|---|---|
| **MANAGERS** | | | | |
| All | -1.51 | -0.26 | 0.81 | 0.44 |
| Male | -2.25 | -0.21 | 0.80 | 1.24 |
| Female | -0.81 | -0.31 | 0.82 | -0.32 |
| **PROFESSIONALS** | | | | |
| All | -1.01 | -0.28 | 0.74 | -0.01 |
| Male | -1.51 | -0.09 | 0.61 | 0.82 |
| Female | -0.83 | -0.35 | 0.79 | -0.31 |
| **TECHNICIANS AND ASSOCIATE PROFESSIONALS** | | | | |
| All | -0.01 | -0.09 | 0.13 | -0.22 |
| Male | -0.19 | -0.07 | 0.08 | 0.04 |
| Female | 0.06 | -0.11 | 0.16 | -0.33 |
| **CLERICAL SUPPORT WORKERS** | | | | |
| All | 0.45 | 0.01 | -0.29 | -0.15 |
| Male | 0.16 | 0.66 | -0.01 | 0.51 |
| Female | 0.49 | -0.09 | -0.33 | -0.24 |
| **SERVICE AND SALES WORKERS** | | | | |
| All | 0.50 | 0.01 | -0.27 | -0.22 |
| Male | -0.77 | -0.26 | 0.09 | 0.42 |
| Female | 0.92 | 0.10 | -0.38 | -0.43 |
| **CRAFT AND RELATED TRADES WORKERS** | | | | |
| All | 0.79 | 0.39 | -0.49 | 0.08 |
| Male | 0.57 | 0.39 | -0.43 | 0.25 |
| Female | 1.33 | 0.37 | -0.64 | -0.32 |
| **PLANT AND MACHINE OPERATORS AND ASSEMBLERS** | | | | |
| All | 1.13 | 0.89 | -0.69 | 0.45 |
| Male | 0.27 | 0.56 | -0.57 | 0.87 |
| Female | 2.49 | 1.41 | -0.88 | -0.20 |
| **ELEMENTARY OCCUPATIONS** | | | | |
| All | 1.16 | -0.25 | -1.08 | -0.33 |
| Male | 0.82 | -0.35 | -1.06 | -0.11 |
| Female | 1.35 | -0.20 | -1.09 | -0.46 |

Source: Authors' elaboration/calculation.

# TABLE A12

## VIETNAM—TASK MEASURES OF MAJOR OCCUPATION GROUPS BY GENDER

| | RTI index | Routine task index | Abstract task index | Manual task index |
|---|---|---|---|---|
| **MANAGERS** | | | | |
| All | -1.90 | 0.03 | 1.16 | 0.77 |
| Male | -2.24 | -0.08 | 1.17 | 0.99 |
| Female | -1.15 | 0.26 | 1.14 | 0.28 |
| **PROFESSIONALS** | | | | |
| All | -0.88 | 0.07 | 0.84 | 0.11 |
| Male | -1.15 | 0.25 | 0.93 | 0.48 |
| Female | -0.72 | -0.05 | 0.79 | -0.12 |
| **TECHNICIANS AND ASSOCIATE PROFESSIONALS** | | | | |
| All | -0.71 | -0.04 | 0.53 | 0.14 |
| Male | -1.19 | 0.06 | 0.75 | 0.50 |
| Female | -0.42 | -0.10 | 0.40 | -0.07 |
| **CLERICAL SUPPORT WORKERS** | | | | |
| All | -0.13 | -0.01 | 0.19 | -0.07 |
| Male | -0.52 | -0.09 | 0.10 | 0.33 |
| Female | 0.08 | 0.03 | 0.24 | -0.29 |
| **SERVICE AND SALES WORKERS** | | | | |
| All | 0.11 | -0.24 | -0.16 | -0.19 |
| Male | -0.30 | -0.25 | -0.01 | 0.06 |
| Female | 0.33 | -0.23 | -0.24 | -0.33 |
| **SKILLED AGRICULTURAL, FORESTRY AND FISHERY WORKERS** | | | | |
| All | 0.46 | 0.05 | -0.53 | 0.12 |
| Male | 0.48 | 0.03 | -0.40 | -0.05 |
| Female | 0.41 | 0.12 | -1.01 | 0.72 |
| **CRAFT AND RELATED TRADES WORKERS** | | | | |
| All | 0.70 | 0.45 | -0.28 | 0.03 |
| Male | 0.13 | 0.49 | -0.01 | 0.37 |
| Female | 1.28 | 0.41 | -0.55 | -0.32 |
| **PLANT AND MACHINE OPERATORS AND ASSEMBLERS** | | | | |
| All | 0.13 | 0.55 | -0.50 | 0.91 |
| Male | -0.39 | 0.45 | -0.33 | 1.17 |
| Female | 1.85 | 0.86 | -1.06 | 0.07 |
| **ELEMENTARY OCCUPATIONS** | | | | |
| All | 0.92 | -0.16 | -0.79 | -0.29 |
| Male | 0.81 | -0.13 | -0.74 | -0.19 |
| Female | 0.98 | -0.17 | -0.81 | -0.34 |

Source: Authors' elaboration/calculation.

## TABLE A13

### YUNNAN (CHINA)—TASK MEASURES OF MAJOR OCCUPATION GROUPS BY GENDER

| | RTI index | Routine task index | Abstract task index | Manual task index |
|---|---|---|---|---|
| **MANAGERS** All | -1.22 | -0.18 | 0.69 | 0.35 |
| Male | -1.32 | -0.12 | 0.71 | 0.49 |
| Female | -1.05 | -0.27 | 0.65 | 0.13 |
| **PROFESSIONALS** All | -0.58 | -0.03 | 0.57 | -0.02 |
| Male | -1.20 | -0.15 | 0.72 | 0.33 |
| Female | -0.11 | 0.07 | 0.46 | -0.29 |
| **TECHNICIANS AND ASSOCIATE PROFESSIONALS** All | -0.52 | -0.12 | 0.43 | -0.03 |
| Male | -0.90 | -0.20 | 0.42 | 0.28 |
| Female | -0.08 | -0.02 | 0.45 | -0.39 |
| **CLERICAL SUPPORT WORKERS** All | 0.40 | 0.11 | -0.11 | -0.18 |
| Male | 0.11 | 0.32 | 0.02 | 0.20 |
| Female | 0.56 | -0.01 | -0.18 | -0.40 |
| **SERVICE AND SALES WORKERS** | -0.03 | -0.09 | -0.05 | -0.02 |
| All | -0.34 | -0.13 | 0.02 | 0.19 |
| Male | 0.34 | -0.05 | -0.12 | -0.26 |
| Female | | | | |
| **SKILLED AGRICULTURAL, FORESTRY AND FISHERY WORKERS** | -0.08 | -0.88 | -0.85 | 0.05 |
| All | -0.68 | -1.02 | -0.75 | 0.40 |
| Male | 0.98 | -0.63 | -1.04 | -0.57 |
| Female | | | | |
| **CRAFT AND RELATED TRADES WORKERS** | 1.12 | 0.84 | -0.34 | 0.07 |
| All | 1.10 | 1.01 | -0.27 | 0.18 |
| Male | 1.17 | 0.33 | -0.55 | -0.29 |
| Female | | | | |
| **PLANT AND MACHINE OPERATORS AND ASSEMBLERS** | -0.02 | 0.30 | -0.41 | 0.73 |
| All | -0.07 | 0.32 | -0.48 | 0.87 |
| Male | 0.15 | 0.24 | -0.20 | 0.29 |
| Female | | | | |
| **ELEMENTARY OCCUPATIONS** | 0.46 | -0.27 | -0.43 | -0.30 |
| All | -0.08 | -0.27 | -0.10 | -0.09 |
| Male | 1.06 | -0.27 | -0.80 | -0.53 |
| Female | | | | |

Source: Authors' elaboration/calculation.

## TABLE A14
### ARMENIA - OLS REGRESSIONS OF RTI INDEX

| | (1) | (2) | (3) | (4) | (5) |
|---|---|---|---|---|---|
| Female | 0.62***<br>(0.15) | 0.74***<br>(0.14) | 0.74***<br>(0.14) | 0.74***<br>(0.14) | 0.83***<br>(0.15) |
| Less than high-school | | 0.90***<br>(0.31) | 0.82***<br>(0.31) | 0.81**<br>(0.31) | 0.23<br>(0.31) |
| More than high-school | | -0.69***<br>(0.13) | -0.70***<br>(0.13) | -0.69***<br>(0.13) | -0.13<br>(0.12) |
| Experience | | | -0.01<br>(0.02) | -0.01<br>(0.02) | -0.03**<br>(0.02) |
| Experience squared | | | 0.00**<br>(0.00) | 0.00**<br>(0.00) | 0.00***<br>(0.00) |
| Bilingual or non-native speaker | | | | -0.19<br>(0.16) | -0.07<br>(0.15) |
| 2-digit occupation | No | No | No | No | Yes |
| Mean RTI | 0.00 | 0.00 | -0.00 | -0.00 | -0.00 |
| N | 989.00 | 988.00 | 972.00 | 972.00 | 972.00 |
| R-squared | 0.03 | 0.08 | 0.10 | 0.10 | 0.28 |

Source: Authors' elaboration/calculation.
Notes: Standard errors are in parentheses. All models include a constant and simultaneously control for weighting, clustering, and stratification. Native speaking male high-school graduates are the reference group. * p < 0.1, ** p < 0.05, *** p < 0.01.

## TABLE A15
### BOLIVIA - OLS REGRESSIONS OF RTI INDEX

| | (1) | (2) | (3) | (4) | (5) |
|---|---|---|---|---|---|
| Female | 0.81*** (0.12) | 0.76*** (0.11) | 0.78*** (0.12) | 0.78*** (0.12) | 0.54*** (0.12) |
| Less than high-school | | 0.04 (0.13) | 0.08 (0.14) | 0.09 (0.14) | -0.03 (0.14) |
| More than high-school | | -1.04*** (0.14) | -0.99*** (0.14) | -0.99*** (0.14) | -0.51*** (0.16) |
| Experience | | | -0.04** (0.02) | -0.04** (0.02) | -0.03* (0.01) |
| Experience squared | | | 0.00** (0.00) | 0.00** (0.00) | 0.00* (0.00) |
| Bilingual or non-native speaker | | | | -0.05 (0.12) | -0.09 (0.12) |
| 2-digit occupation | No | No | No | No | Yes |
| Mean RTI | -0.00 | -0.01 | -0.01 | -0.01 | -0.01 |
| N | 1757.00 | 1745.00 | 1745.00 | 1735.00 | 1735.00 |
| R-squared | 0.05 | 0.14 | 0.15 | 0.15 | 0.34 |

Source: Authors' elaboration/calculation.
Notes: Standard errors are in parentheses. All models include a constant and simultaneously control for weighting, clustering, and stratification. Native speaking male high-school graduates are the reference group. * $p < 0.1$, ** $p < 0.05$, *** $p < 0.01$.

## TABLE A16

### COLOMBIA - OLS REGRESSIONS OF RTI INDEX

| | (1) | (2) | (3) | (4) | (5) |
|---|---|---|---|---|---|
| Female | 0.41***<br>(0.06) | 0.41***<br>(0.08) | 0.41***<br>(0.08) | 0.42***<br>(0.08) | 0.21<br>(0.13) |
| Less than high-school | | -0.07<br>(0.07) | 0.06<br>(0.08) | 0.02<br>(0.09) | 0.04<br>(0.08) |
| More than high-school | | -0.84***<br>(0.06) | -0.83***<br>(0.07) | -0.83***<br>(0.07) | -0.31**<br>(0.10) |
| Experience | | | -0.03*<br>(0.02) | -0.04*<br>(0.02) | -0.03*<br>(0.01) |
| Experience squared | | | 0.00<br>(0.00) | 0.00*<br>(0.00) | 0.00<br>(0.00) |
| Bilingual or non-native speaker | | | | 1.99***<br>(0.19) | 1.66***<br>(0.47) |
| 2-digit occupation | No | No | No | No | Yes |
| Mean RTI | -0.00 | -0.01 | -0.01 | -0.01 | -0.01 |
| N | 1716.00 | 1704.00 | 1704.00 | 1704.00 | 1704.00 |
| R-squared | 0.01 | 0.06 | 0.07 | 0.07 | 0.28 |

Source: Authors' elaboration/calculation.
Notes: Standard errors are in parentheses. All models include a constant and simultaneously control for weighting, clustering, and stratification. Native speaking male high-school graduates are the reference group. * $p < 0.1$, ** $p < 0.05$, *** $p < 0.01$.

## TABLE A17

### GEORGIA - OLS REGRESSIONS OF RTI INDEX

| | (1) | (2) | (3) | (4) | (5) |
|---|---|---|---|---|---|
| Female | 0.42***<br>(0.15) | 0.51***<br>(0.15) | 0.49***<br>(0.15) | 0.49***<br>(0.15) | 0.67***<br>(0.16) |
| Less than high-school | | 0.36<br>(0.34) | 0.37<br>(0.34) | 0.32<br>(0.34) | -0.22<br>(0.34) |
| More than high-school | | -0.66***<br>(0.16) | -0.69***<br>(0.16) | -0.68***<br>(0.16) | -0.31*<br>(0.18) |
| Experience | | | 0.03**<br>(0.02) | 0.03*<br>(0.02) | 0.03<br>(0.02) |
| Experience squared | | | -0.00<br>(0.00) | -0.00<br>(0.00) | -0.00<br>(0.00) |
| Bilingual or non-native speaker | | | | 0.19<br>(0.27) | 0.18<br>(0.27) |
| 2-digit occupation | No | No | No | No | Yes |
| Mean RTI | -0.00 | 0.00 | -0.00 | -0.00 | -0.00 |
| N | 933.00 | 933.00 | 932.00 | 932.00 | 932.00 |
| R-squared | 0.01 | 0.04 | 0.05 | 0.05 | 0.22 |

Source: Authors' elaboration/calculation.
Notes: Standard errors are in parentheses. All models include a constant and simultaneously control for weighting and clustering. Native speaking male high-school graduates are the reference group. * $p < 0.1$, ** $p < 0.05$, *** $p < 0.01$.

## TABLE A18
### GHANA - OLS REGRESSIONS OF RTI INDEX

| | (1) | (2) | (3) | (4) | (5) |
|---|---|---|---|---|---|
| Female | 0.85*** (0.09) | 0.77*** (0.10) | 0.78*** (0.10) | 0.78*** (0.10) | 0.44*** (0.10) |
| Less than high-school | | 0.23 (0.14) | 0.40*** (0.14) | 0.40*** (0.14) | 0.08 (0.14) |
| More than high-school | | -0.98*** (0.19) | -0.92*** (0.18) | -0.92*** (0.18) | -0.67*** (0.18) |
| Experience | | | -0.05*** (0.01) | -0.05*** (0.01) | -0.03*** (0.01) |
| Experience squared | | | 0.00*** (0.00) | 0.00*** (0.00) | 0.00*** (0.00) |
| Bilingual or non-native speaker | | | | -0.01 (0.10) | -0.01 (0.08) |
| 2-digit occupation | No | No | No | No | Yes |
| Mean RTI | 0.00 | -0.05 | -0.05 | -0.05 | -0.05 |
| N | 2133.00 | 1895.00 | 1895.00 | 1892.00 | 1892.00 |
| R-squared | 0.07 | 0.12 | 0.13 | 0.13 | 0.31 |

Source: Authors' elaboration/calculation.
Notes: Standard errors are in parentheses. All models include a constant and simultaneously control for weighting, clustering, and stratification. Native speaking male high-school graduates are the reference group. * p < 0.1, ** p < 0.05, *** p < 0.01.

## TABLE A19

### KENYA - OLS REGRESSIONS OF RTI INDEX

| | (1) | (2) | (3) | (4) | (5) |
|---|---|---|---|---|---|
| Female | 0.27***<br>(0.08) | 0.25***<br>(0.08) | 0.23***<br>(0.08) | 0.24***<br>(0.08) | 0.21***<br>(0.08) |
| Less than high-school | | 0.46***<br>(0.10) | 0.56***<br>(0.10) | 0.55***<br>(0.10) | 0.35***<br>(0.10) |
| More than high-school | | -0.91***<br>(0.13) | -0.94***<br>(0.13) | -0.94***<br>(0.13) | -0.48***<br>(0.13) |
| Experience | | | -0.04***<br>(0.01) | -0.04***<br>(0.01) | -0.05***<br>(0.01) |
| Experience squared | | | 0.00**<br>(0.00) | 0.00**<br>(0.00) | 0.00***<br>(0.00) |
| Bilingual or non-native speaker | | | | 0.11<br>(0.11) | 0.13<br>(0.11) |
| 2-digit occupation | No | No | No | No | Yes |
| Mean RTI | 0.00 | -0.07 | -0.06 | -0.06 | -0.06 |
| N | 2361.00 | 2150.00 | 2134.00 | 2131.00 | 2131.00 |
| R-squared | 0.01 | 0.10 | 0.11 | 0.11 | 0.28 |

Source: Authors' elaboration/calculation.
Notes: Standard errors are in parentheses. All models include a constant and simultaneously control for weighting, clustering, and stratification. Native speaking male high-school graduates are the reference group. * $p < 0.1$, ** $p < 0.05$, *** $p < 0.01$.

## TABLE A20

### LAOS - OLS REGRESSIONS OF RTI INDEX

| | (1) | (2) | (3) | (4) | (5) |
|---|---|---|---|---|---|
| Female | 0.36*** (0.12) | 0.32** (0.13) | 0.23* (0.12) | 0.25* (0.13) | 0.14 (0.12) |
| Less than high-school | | 0.46** (0.18) | 0.75*** (0.19) | 0.75*** (0.19) | 0.42*** (0.15) |
| More than high-school | | -0.85*** (0.23) | -0.92*** (0.22) | -0.90*** (0.22) | -0.36* (0.19) |
| Experience | | | -0.05*** (0.02) | -0.05*** (0.02) | -0.04*** (0.01) |
| Experience squared | | | 0.00* (0.00) | 0.00* (0.00) | 0.00 (0.00) |
| Bilingual or non-native speaker | | | | 0.11 (0.16) | -0.05 (0.13) |
| 2-digit occupation | No | No | No | No | Yes |
| Mean RTI | -0.00 | -0.05 | -0.05 | -0.05 | -0.05 |
| N | 2185.00 | 2004.00 | 2004.00 | 2004.00 | 2004.00 |
| R-squared | 0.01 | 0.10 | 0.13 | 0.13 | 0.28 |

Source: Authors' elaboration/calculation.
Notes: Standard errors are in parentheses. All models include a constant and simultaneously control for weighting, clustering, and stratification. Native speaking male high-school graduates are the reference group. * $p < 0.1$, ** $p < 0.05$, *** $p < 0.01$.

# TABLE A21
## MACEDONIA - OLS REGRESSIONS OF RTI INDEX

| | (1) | (2) | (3) | (4) | (5) |
|---|---|---|---|---|---|
| Female | 0.60***<br>(0.11) | 0.75***<br>(0.10) | 0.77***<br>(0.10) | 0.74***<br>(0.10) | 0.66***<br>(0.10) |
| Less than high-school | | 0.84***<br>(0.17) | 0.79***<br>(0.17) | 0.88***<br>(0.17) | 0.32**<br>(0.15) |
| More than high-school | | -1.33***<br>(0.10) | -1.38***<br>(0.10) | -1.38***<br>(0.10) | -0.60***<br>(0.12) |
| Experience | | | -0.05***<br>(0.02) | -0.05***<br>(0.02) | -0.05***<br>(0.01) |
| Experience squared | | | 0.00***<br>(0.00) | 0.00***<br>(0.00) | 0.00***<br>(0.00) |
| Bilingual or non-native speaker | | | | -0.36***<br>(0.14) | -0.21*<br>(0.12) |
| 2-digit occupation | No | No | No | No | Yes |
| Mean RTI | 0.00 | -0.00 | -0.00 | 0.00 | 0.00 |
| N | 1810.00 | 1809.00 | 1809.00 | 1808.00 | 1808.00 |
| R-squared | 0.02 | 0.17 | 0.18 | 0.18 | 0.34 |

Source: Authors' elaboration/calculation.
Notes: Standard errors are in parentheses. All models include a constant and simultaneously control for weighting and clustering. Native speaking male high-school graduates are the reference group. * $p < 0.1$, ** $p < 0.05$, *** $p < 0.01$

## TABLE A22

### SRI LANKA - OLS REGRESSIONS OF RTI INDEX

| | (1) | (2) | (3) | (4) | (5) |
|---|---|---|---|---|---|
| Female | 0.59***<br>(0.10) | 0.69***<br>(0.09) | 0.70***<br>(0.09) | 0.70***<br>(0.09) | 0.53***<br>(0.10) |
| Less than high-school | | 0.83***<br>(0.11) | 0.77***<br>(0.13) | 0.77***<br>(0.13) | 0.38***<br>(0.13) |
| More than high-school | | -0.52***<br>(0.17) | -0.51***<br>(0.17) | -0.50***<br>(0.16) | -0.33**<br>(0.15) |
| Experience | | | -0.02<br>(0.02) | -0.02<br>(0.02) | -0.01<br>(0.02) |
| Experience squared | | | 0.00<br>(0.00) | 0.00<br>(0.00) | 0.00<br>(0.00) |
| Bilingual or non-native speaker | | | | 0.06<br>(0.15) | 0.01<br>(0.14) |
| 2-digit occupation | No | No | No | No | Yes |
| Mean RTI | 0.00 | -0.01 | -0.01 | -0.01 | -0.01 |
| N | 1559.00 | 1545.00 | 1543.00 | 1540.00 | 1540.00 |
| R-squared | 0.03 | 0.13 | 0.13 | 0.13 | 0.27 |

Source: Authors' elaboration/calculation.
Notes: Standard errors are in parentheses. All models include a constant and simultaneously control for weighting, clustering, and stratification. Native speaking male high-school graduates are the reference group. * $p < 0.1$, ** $p < 0.05$, *** $p < 0.01$.

## TABLE A23

### VIETNAM - OLS REGRESSIONS OF RTI INDEX

| | (1) | (2) | (3) | (4) | (5) |
|---|---|---|---|---|---|
| Female | 0.68*** (0.08) | 0.68*** (0.08) | 0.67*** (0.08) | 0.68*** (0.08) | 0.54*** (0.08) |
| Less than high-school | | 0.56*** (0.10) | 0.60*** (0.10) | 0.58*** (0.11) | 0.28*** (0.10) |
| More than high-school | | -0.86*** (0.11) | -0.90*** (0.12) | -0.90*** (0.12) | -0.32*** (0.12) |
| Experience | | | -0.03** (0.01) | -0.03** (0.01) | -0.02** (0.01) |
| Experience squared | | | 0.00** (0.00) | 0.00** (0.00) | 0.00** (0.00) |
| Bilingual or non-native speaker | | | | 0.15 (0.15) | 0.28** (0.12) |
| 2-digit occupation | No | No | No | No | Yes |
| Mean RTI | 0.00 | -0.00 | -0.00 | -0.00 | -0.00 |
| N | 2332.00 | 2321.00 | 2321.00 | 2319.00 | 2319.00 |
| R-squared | 0.04 | 0.15 | 0.15 | 0.15 | 0.29 |

Source: Authors' elaboration/calculation.
Notes: Standard errors are in parentheses. All models include a constant and simultaneously control for weighting, clustering, and stratification. Native speaking male high-school graduates are the reference group.
* p < 0.1, ** p < 0.05, *** p < 0.01.

## TABLE A24
### YUNNAN PROVINCE (CHINA) - OLS REGRESSIONS OF RTI INDEX

| | (1) | (2) | (3) | (4) | (5) |
|---|---|---|---|---|---|
| Female | 0.64*** (0.10) | 0.72*** (0.10) | 0.81*** (0.10) | 0.80*** (0.10) | 0.73*** (0.11) |
| Less than high-school | | 0.81*** (0.10) | 0.69*** (0.12) | 0.68*** (0.12) | 0.48*** (0.12) |
| More than high-school | | -0.37 (0.89) | -0.43 (0.80) | -0.47 (0.81) | -0.40 (0.74) |
| Experience | | | -0.08*** (0.01) | -0.08*** (0.01) | -0.08*** (0.01) |
| Experience squared | | | 0.00*** (0.00) | 0.00*** (0.00) | 0.00*** (0.00) |
| Bilingual or non-native speaker | | | | -0.25 (0.17) | -0.20 (0.15) |
| 2-digit occupation | No | No | No | No | Yes |
| Mean RTI | 0.00 | -0.01 | -0.01 | -0.01 | -0.01 |
| N | 1244.00 | 1238.00 | 1238.00 | 1238.00 | 1238.00 |
| R-squared | 0.03 | 0.08 | 0.12 | 0.12 | 0.24 |

Source: Authors' elaboration/calculation.
Notes: Standard errors are in parentheses. All models include a constant and simultaneously control for weighting and clustering. Native speaking male high-school graduates are the reference group. * $p < 0.1$, ** $p < 0.05$, *** $p < 0.01$.